\documentclass{revtex4}
\usepackage{graphicx}
\usepackage{epsfig}
\usepackage{amsmath}
\usepackage{amssymb}

\newcommand{\ee}{\end{eqnarray}}
\newcommand{\be}[1]{\begin{eqnarray} \mbox{$\label{#1}$} }
\newcommand{\mtrx}[2]{\left(\begin{array}{#1} #2 \end{array}\right)}
\newcommand{\eref}[1]{(\ref{#1})}
\newcommand{\der}{\mathrm{d}}
\newcommand{\tr}{\mathrm{Tr}\,}

\newcommand\ie {{\it i.e.~}}
\newcommand\eg {{\it e.g.~}}

\newcommand\etal{{\it et al.~}}

\newcommand{\ul}\underline

\newcommand{\GR}{G^\mathrm{R}}
\newcommand{\SR}{\Sigma^\mathrm{R}}
\newcommand{\GoR}{G^{0\mathrm{R}} }
\newcommand{\GA}{G^\mathrm{A}}
\newcommand{\SA}{\Sigma^\mathrm{A}}
\newcommand{\GoA}{G^{0\mathrm{A}} }

\newcommand{\GL}{G^<}
\newcommand{\SL}{\Sigma^<}

\newcommand{\prip}[1]{\mathcal{P}\left(#1\right)}
\newcommand{\pripp}{\mathcal{P}_+}
\newcommand{\pripm}{\mathcal{P}_-}
\newcommand{\prippm}{\mathcal{P}_\pm}

\newcommand{\evu}{\mathcal{U}}

\newcommand{\DT}{\Delta}
\newcommand{\pr}{^\prime}

\renewcommand{\vec}[1]{\text{\boldmath{$ #1 $}}}
\newcommand{\uv}[1]{\vec{\hat{#1}}} 
\newcommand{\ksum}[1]{\int_{#1}}  
\newcommand{\ki}{\ksum{\vec{k\pr}}}

\newcommand{\Wkk}{W_{\vec{k}\vec{k\pr}}}

\newcommand{\fermi}{\mathrm{F}}

\newcommand{\sch}{Schr\"odinger{ }}

\newcommand{\feq}{f^\textrm{eq}}
\newcommand{\fee}{f^{(E)}}

\newcommand{\erg}{\epsilon}

\newcommand{\ben}{b}
\newcommand{\cen}{c}
\newcommand{\sproj}{S}
\newcommand{\idm}{\mathbf{1}}
\newcommand{\ttr}{\tau_\textrm{tr}}

\newcommand{\tTh}{\tau_{\textrm{Th}}}
\newcommand{\gT}{g_\mathrm{T}}
\newcommand{\cD}{c_i}

\newcommand{\vkm}{\mathbf{v}}

\newcommand{\ordo}[1]{\mathcal{O}(#1)}
\newcommand{\FD}[1]{f_\mathrm{FD}\left(#1\right)}
\newcommand{\FDf}{f_\mathrm{FD}}

\newcommand{\DS}{D} 

\newcommand{\JC}{\mathcal{J}}
\newcommand{\JCd}{\mathcal{J}^\delta}
\newcommand{\JCp}{\mathcal{J}^\mathrm{P}}
\newcommand{\JCpx}{\mathcal{J}^{\mathrm{P}X}}
\newcommand{\JCpy}{\mathcal{J}^{\mathrm{P}Y}}
\newcommand{\IC}{\mathcal{I}}

\renewcommand{\Re}{\mathrm{Re}\,}
\renewcommand{\Im}{\mathrm{Im}\,}

\newcommand{\GO}{G1}
\newcommand{\GT}{G2}
\renewcommand{\AA}{AA}
\newcommand{\SKBA}{SKBA}

\newcommand{\Htot}{H_\textrm{tot}}
\newcommand{\Ho}{H_0}
\newcommand{\Hi}{V}%{H_\textrm{i}}
\newcommand{\rhoI}{\rho^{\textrm{I}}} 
\newcommand{\INT}{^\textrm{I}} 

\newcommand{\nimp}{n_\textrm{imp}}
\newcommand{\singleimp}{u}

\newcommand{\cH}{\gamma}
\newcommand{\cp}{\kappa} 

\newcommand{\Drude}{\sigma_{\textrm{D}}}

\newcommand{\Ndisp}{{N_\mathrm{d}}}
\newcommand{\Nchiral}{{N_\mathrm{c}}}

\newcommand{\Ansatzindex}{\nu}

\begin{document}

\title{Theory of Conductivity of Chiral Particles}

\author{Janik Kailasvuori}

\affiliation{International Institute of Physics, Universidade Federal do Rio Grande do Norte,
59078-400 Natal-RN,
Brazil}

\affiliation{Max-Planck-Institut f\"ur Physik komplexer Systeme, N\"othnitzer Str. 38, 01187 Dresden, Germany}

\author{B{\v r}etislav {\v S}op{\' i}k}

\affiliation{Central European Institute of Technology, Masaryk University, 
Kamenice 735, 62500 Brno, Czech Republic}

\author{Maxim Trushin}

\affiliation{University of Konstanz, Fachbereich Physik M703, 78457 Konstanz, Germany}

\begin{abstract}
\noindent
In this methodology focused paper we scrutinize the application of  the band-coherent Boltzmann equation approach to  calculating the conductivity of chiral particles. As the ideal testing ground we use the two-band kinetic Hamiltonian with an $N$-fold chiral twist that arise in a low-energy description of charge carriers in rhombohedrally stacked multilayer graphene. To understand the role of chirality in the conductivity of such particles we also consider the artificial model with the chiral winding number decoupled from the power of the dispersion. We first utilize the approximate but analytically solvable band-coherent Boltzmann approach including the ill-understood principal value terms that are a byproduct of several quantum-many body theory derivations of Boltzmann collision integrals. Further on, we employ the finite-size Kubo formula with the exact diagonalization of the total Hamiltonian perturbed by disorder. Finally, we compare several choices of {\it Ansatz} in the derivation of the Boltzmann equation according to the qualitative agreement between the Boltzmann and Kubo conductivities. We find that the best agreement can be reached in the approach where the 
principle value terms in the collision integral are absent.
\end{abstract}

\date{\today}
\maketitle

\section{Introduction}
\noindent
The isolation of the one-atomic carbon layers \cite{Science2004novoselov} has boosted an extremely intensive research
in the field of quantum transport associated with Dirac electrons.\cite{RMP2009castroneto}
One of the primary reasons of such an overwhelmed interest is the bizarre chiral nature of the Dirac carriers
which makes the conduction and valence states entangled with each other.
This coupling leads to many nontrivial effects,\cite{RMP2009castroneto} including the famous Klein tunneling \cite{Nature2006katsnelson}. 

The chiral carriers are usually characterized by means of the pseudospin which is an additional quantum number
formally similar to the real spin in spin-orbit coupled systems. 
In a single layer graphene, the pseudospin keeps its radial orientation along the Fermi circle 
resulting in the winding number $N=1$. The winding number is directly related to the
Berry phase $\pi N$ responsible for the anomalous quantum Hall effect observed at the very beginning
of the graphene rush.\cite{Novoselov-Geim-etal-2005a, Zhang-Kim-etal-2005a}
There are quite a few speculations on the important role of the pseudospin-coherent contribution in
the electrical dc conductivity of Dirac fermions at low carrier concentrations. 
\cite{Auslender-Katsnelson-2007,Trushin-Schliemann-2007,Kailasvuori-Lueffe-2010,Culcer-Winkler-2007b, Liu-Lei-Horing-2008}
There is however no agreement whether this contribution really matters and can be extracted from the total conductivity.

There are two aspects of this problem worthy of discussion. The first one is experimental:
It is very difficult experimentally to detect the pseudospin-coherent contribution because
the subtle quantum mechanical effects responsible for this contribution are easily smeared out by temperature, ripples,
and charged inhomogeneities inherited in graphene. In spite of the impressive progress\cite{Nano2012mayorov}
made with suspended graphene, this contribution has not been extracted yet.
The second problem is purely theoretical: It is still not clear how large the pseudospin contribution should be
since the numerical methods can give only an estimation \cite{Nomura-MacDonald-2007}  while 
the approximated Boltzmann-like approaches \cite{Auslender-Katsnelson-2007,Trushin-Schliemann-2007,Kailasvuori-Lueffe-2010,Culcer-Winkler-2007b, Liu-Lei-Horing-2008}
are unreliable at low carrier concentrations and may lead to different answers depending on the approximations involved.

The main goal of the present work is to compare the exact numerical (finite-size Kubo) and approximated analytical (Boltzmann-like)
pseudospin-coherent conductivities and in that way to figure out the correct approximation needed to derive
the most reasonable analytical expression. In particular, we focus on the dependence of pseudospin-coherent conductivity
on the winding number $N$. Surprisingly, the semiclassical pseudospin-coherent conductivity acquires qualitatively different
behavior on $N$ depending on the approximation done. The subsequent comparison with the numerical calculation
helps us to rule out the ``wrong'' approximations and keep the ``correct'' ones. 

Note, that the chiral particles with an arbitrary winding number
can be  realized in the chirally stacked multilayer graphene.
The recent investigations of the band structure in a few layer graphene \cite{PRL2006latil,PRB2006guinea,PRB2010asvetisyan} 
indicate its strong sensitivity to the stacking pattern.
Graphene layers placed together in the most natural way do not lie exactly one on top of the other
but are shifted in such a way that only half of the carbon atoms
have a neighbor in another layer and the other half are projected
right into the middle of the graphene's honeycomb cell.
If the third layer aligns with the first (and the $N+2$ layer with the $N$-th) then we arrive at
the most abundantly occurring arrangement of graphene layers known as Bernal (or ABAB...) stacking.
An alternative stacking is obtained when the third layer aligns
neither with the first nor with the second but is shifted with respect to both, and it is the fourth layer that again aligns with the first.
This arrangement is known as rhombohedral (or ABCABC...) stacking. The simplest model used in the study of the latter system \cite{Min-MacDonald-2008} 
is the model that we shall use as our testing ground in the investigation on theoretical methods to calculate the conductivity. 

The beauty of the effective low-energy Hamiltonian describing the carriers in the chirally stacked multilayer graphene has attracted
many theoreticians working in the field of quantum transport. Among the most recent works are 
the theory of chirality-dependent phonon scattering \cite{PRB2011amin},
the semiclassical Boltzmann transport theory \cite{PRB2011bmin},
the investigations of Landau levels\cite{PRB2011koshino},
Berry phase \cite{PRB2009koshino}, effects of screening \cite{PRB2010koshino,PRB2012min},
optical conductivity\cite{PRL2009min}, and THz absorption \cite{NJP2012trushin}.
We would like to warn the reader, that although our ultimate interest is to understand conductivity in graphene type of systems, the aim of this paper is purely methodological,
and we undertake no discussion about the connection to realistic systems and to experimental findings.
For example, we will for simplicity and analytical solvability assume a fully homogenous background, whereas realistic graphene might have to be modeled 
with a variable potential to account for ripples and other sources of inhomogeneity.

The outline of the paper is the following.
 In section~\ref{s:prel} we discuss some preliminaries that will be used in later sections. Section~\ref{s:background} offers in more detail some of the background and motivation of the present paper. In section~\ref{s:kuboheuristics} we present an analytical heuristic derivation of the conductivity by evaluating the Kubo formula using the chiral plane-wave states of free electrons of multilayer graphene. The derivation, interesting in its own right, will serve among other as a easy introduction to band-coherent effects and to the distinction between intra-band and inter-band contributions to the conductivity. In section~\ref{s:boltzmann} we will discuss the conductivity derived with a band-coherent Boltzmann equation approach. In section~\ref{s:num} we present the results of the  numerical studies of the Kubo formula.
In section~\ref{s:conclusion} we wrap up with a discussion.

\section{Preliminaries}
\label{s:prel}
\noindent 
Many properties of bilayer graphene can be well understood in terms of two independent, time-reversal related copies of  the effective two-band Hamiltonian\cite{McCann-Falko-2006} 
\be{bilayer}
H_0 = \frac{1}{2m^*}  \mtrx{cc}
{ 0 & (p_x-ip_y )^2 \\
(p_x+i p_y)^2 & 0 } \, .
\ee
This hamiltonian exhibits a coupling between momentum and a spin type of variable, a pseudo-spin that in this case derives from the non-equivalent lattice sites, one from each layer. 
 For $N$-layer rhombohedrally stacked graphene one can with similar approximations  derive an effective low-energy hamiltonian that  also turns also out to be a two band model, this time given by \cite{Min-MacDonald-2008}
\be{multilayer}
H_0 = \cH  \mtrx{cc}
{ 0 & (p_x-ip_y )^N \\
(p_x+i p_y)^N & 0 } \, ,
\ee
with $\cH$ (like $m^*$ above) being determined by material parameters of the system. 
As  a realistic models for rhombohedral multilayer graphene the hamiltonian \eref{multilayer} is expected to become increasingly poor with an increasing number of layers. We  wish to exploit the hamiltonian \eref{multilayer} for theoretical and methodological studies because of its combination of  simplicity and yet very  non-trivial chiral properties. 
For our studies it will turn out to be interesting to generalize the hamiltonian \eref{multilayer} to the more general although artificial hamiltonian   
\be{NdNc}
H_0 =  \vec{\ben} \cdot \vec{\sigma} =  \ben (k) \uv{\ben} (\uv{k})\cdot \vec{\sigma}  = \cH (\hbar k)^{\Ndisp} 
  \mtrx{cc}
{ 0 & e^{-i \Nchiral \theta}  \\
e^{i \Nchiral \theta}   & 0 } \, ,
\ee
where $k=|\vec{k}|$ is the absolute value of the wave number $\vec{k}=k \uv{k}=\vec{p}/\hbar$,  $\theta$ is the angle of the direction $\uv{k}=\vec{k}/k=(\cos \theta, \sin \theta)$ of the wave number and $\vec{\ben} (\vec{k})$ is the (pseudo)spin-orbit field. In the hamiltonian \eref{NdNc}   the power $\Ndisp$ of $k$ in the dispersion $\erg_\vec{k} =\ben=\cH (\hbar k)^\Ndisp$ has been decoupled from the winding number $\Nchiral$ of the direction $\uv{\ben}$ of the pseudospin-orbit field as a function of $\uv{k}$, which determines the chiral properties of the hamiltonian. In the special case  $\Ndisp=\Nchiral$ this hamiltonian reduces to the hamiltonian \eref{multilayer}. The benefit of introducing such an artificial Hamiltonian is that one can take advantage of  some special features of a quadratic dispersion, to be explained later, but still keep the chiral properties general. 

For both the numerical and analytical treatments of the conductivity we will need the velocity matrix, defined as $\vkm_i = \tfrac{\partial H_0}{\partial p_i}$. For the multilayer Hamiltonian \eref{multilayer} one has in the basis of the sublattice orbitals (that is, in the basis in which equation \eref{multilayer} is written) the $x$-component 
\be{noname-06}
\vkm_x =  \cH \hbar^{N-1} \mtrx{cc}{ 0 &  (k_x-i k_y)^{N-1}  \\  (k_x + i k_y)^{N-1} & 0 } \, .
\ee
For us it will sometimes be convenient to express matrix quantities in the eigenbasis of the Hamiltonian, which we also call the chiral basis. For the Hamiltonian \eref{NdNc} the eigenstates are given by 
\be{eigenstates}
\psi_\vec{k}^s  = \frac{ e^{i \vec{k}\cdot \vec{x}} } {L}  \mtrx{c}{ 1\\ s e^{i \Nchiral \theta}} 
\ee   
where $s=\pm $ is the pseudospin index and where $\theta = \arctan (k_y/k_x)$  is the angle of the momentum vector in the momentum plane. The eigenenergies are $\erg_k^s=sb =s \cH (\hbar k)^\Ndisp$. The velocity operator in the chiral basis reads
 \be{noname-01}
\vkm_x = 
\underbrace{ 
\uv{k }_x \hbar^{-1} \partial_k H_0 
}_{
=: \vkm_x^\textrm{intra}
} 
+
\underbrace{ 
\uv{\theta}_x(\hbar k)^{-1} \partial_\theta H_0
}_{
=: \vkm_x^\textrm{inter}
}
=
\hbar^{-1}  \mtrx{cc} {
\uv{k}_x \partial_k\ben  &  i\uv{\theta}_x \frac{\Nchiral \ben}{k}
\\
- i\uv{\theta}\frac{\Nchiral \ben}{k} &- \uv{k}\partial_k\ben  
 }^\textrm{ch} =
 v_k \mtrx{cc} {
\cos \theta &  -i  \frac{\Nchiral }{\Ndisp} \sin\theta
\\
i  \frac{\Nchiral }{\Ndisp} \sin \theta  &- \cos \theta  
 }^\textrm{ch}    
\ee   
where $\uv{\theta}=\uv{z}\times \uv{k}= (-\sin \theta, \cos \theta)$ and $v_k=\partial_k \ben = \Ndisp \cH (\hbar k)^{\Ndisp-1}$. The velocity operator can formally be decomposed into two parts. The intraband part $\vkm_x^\textrm{intra}$ accounts for the ordinary velocity components for each band as where the bands not talking to each other. In the chiral basis it is diagonal.  On the other hand, the interband part $\vkm_x^\textrm{inter}=\uv{\theta}(\hbar k)^{-1} \partial_\theta H_0$ couples the two bands and is an off-diagonal matrix in the chiral basis. The interpretation of these matrix components is  less intuitive.   
They reflect the Zitterbewegung property of Dirac electrons. Both  intraband and interband components of the velocity can contribute to the current. Within the Kubo formula picture we will see this in next section.  In the Boltzmann picture, we note that for  a matrix-valued distribution function $f$ the current in the $x$-direction is given by 
\be{current}
j_x =\int \frac{\der^2 k }{(2\pi)^2} \tr (\vkm_x f  ) =
2 \int_\vec{k}    v_k \left(
\uv{k}_x   f_\uv{\ben}  +  \uv{\theta}_x  f_\uv{\cen}  \frac{\Nchiral}{\Ndisp} 
\right)      
\ee
in terms of the components of the matrix-valued distribution function 
\be{fdecomp}
f=f_0 \idm + \vec{f} \cdot \vec{\sigma}=f_0 \idm + ( f_\uv{\ben} \uv{\ben} + f_\uv{\cen} \uv{\cen} +f_z \uv{z} )\cdot \vec{\sigma} =\mtrx{cc} {
 f_0 + f_\uv{\ben}  & f_z -i f_\uv{\cen}  \\ 
 f_z + i f_\uv{\cen} &   f_0 - f_\uv{\ben} 
 }^\textrm{ch} 
\ee
with the vector part or spin part of $f$ decomposed along the $\theta$-dependent basis vectors  $\{\uv{\ben}, \uv{\cen},\uv{z}\}$ with $\uv{\cen}=\uv{\ben}\times \uv{z}$. The densities $n^\pm =f_0\pm f_\uv{\ben}$ (in phase space) in the upper and lower band, respectively, we find in the chiral basis on the diagonal. Note however, that for the type of Hamiltonians we consider, only the spin component $\vec{f}$ but not the total charge density $2f_0= n^+ +n^- $ contributes to the current. 

The conductivities $\sigma=j_x/E_x$ (the driving electric field is assumed to be in the $x$-direction) will in the present paper be given as functions of the dimensionless variable 
\be{noname-07}
\ell k_\fermi = \ttr v_\fermi k_\fermi =  \Ndisp \frac{\ttr \erg_\fermi}{\hbar}\, .   
 \ee
The momentum relaxation time $\ttr=(\IC (k_\fermi))^{-1} $ is derived from integral
\be{I-int}
\IC (k) = \int \frac{\der^2 k \pr}{(2\pi)^2} \delta (\erg_k-\erg_{k\pr}) \Wkk 
\cos^2 \frac{\Nchiral \Delta \theta}{2} %\frac{1+\cos \Nchiral \Delta \theta}{2} 
(1-\cos \Delta \theta)   
\ee
where $\Wkk = \frac{2\pi }{\hbar} n_i |U_{\vec{k}\vec{k\pr}} L^2|^2$ gives the collision probability rate according to Fermi's golden rule for spinless particles scattered by an impurity potential $U$ and the unusual trigonometric factor $\cos^2 \tfrac{\Nchiral \Delta \theta}{2}$ comes from the spin overlap between the chiral eigenstates at a momentum angle difference  $\Delta\theta = \theta- \theta\pr $. In particular, the Drude conductivity  is given by 
\be{noname-08}
\Drude = \frac{e^2}{h} \cdot  \frac{\ell k_\fermi} {2},
\ee
independently of the dispersion and winding number.

The Kubo formula for the conductivity reads
\be{kubo}
\sigma_\textrm{Kubo} = -i \frac{ \hbar e^2}{L^2} 
 \sum_{n, n\pr}
  \frac{\FD{\erg_n}-\FD{\erg_{n\pr}}}{\erg_n-\erg_{n\pr}} 
   \frac{
   \langle n | \vkm_x | n\pr \rangle  \langle n\pr | \vkm_x | n \rangle 
   }{
   \erg_n - \erg_{n\pr} + i \eta 
   }\, ,
\ee
where $|n\rangle$ are the eigenstates of the total Hamiltonian including impurities but excluding the driving electric potential and $\FD{\erg}$ is the Fermi-Dirac distribution. 
The parameter $\eta$ describes the broadening of the eigenstates. For a pure sample it would be infinitessimal. In our case it is finite, as we  describe a system with impurities. We will find it useful to study the quantities 
\be{sigmaintrainter}
\sigma^\textrm{intra}  & :=&  \sigma_\textrm{Kubo}  (\vkm_x \rightarrow \vkm_x^\textrm{intra} ) 
\nonumber 
\\ 
\sigma^\textrm{inter}  & :=&  \sigma_\textrm{Kubo}  (\vkm_x \rightarrow \vkm_x^\textrm{inter} )\, .
\ee
Numerical checks will show that there is at the most a negligible contribution from the remaining possible terms containing matrix elements $\langle n | \vkm_x^\textrm{intra} | n\pr \rangle  \langle n\pr | \vkm_x^\textrm{inter} | n \rangle$ or $\langle n | \vkm_x^\textrm{inter} | n\pr \rangle  \langle n\pr | \vkm_x^\textrm{intra} | n \rangle$\, .

It is rather clear that the Drude conductivity $\Drude$ as a pure intraband contribution resides completely within  $\sigma^\textrm{intra}$. However, the above definitions of $\sigma^\textrm{intra}$ and $\sigma^\textrm{inter}$ are in general not a clean  decomposition of intraband and interband contributions to the conductivity, wherefore $\sigma^\textrm{intra}$ might on top of $\Drude$ contain some interband contributions. For the problem we study in the present paper the Boltzmann approach gives support for this decomposition faithfully separating pure intraband contributions form interband contributions,  but in other cases, for example for the much more complex case of monolayer graphene ($N=1$), the Boltzmann approach (see ref.~\onlinecite{Kailasvuori-Lueffe-2010}) comes with the nonintuitive message that  there can  be important spin-coherence originated contributions also in the part $f_\uv{\ben}$  of the distribution function that contributes to the current through $\vkm_\textrm{intra}$, although $f_\uv{\ben} =n^+ - n^-$ at a first glance seems to 
be a pure intraband contribution as it describes the polarization along the quantization axis of the Hamiltonian. 

\section{Heuristic evaluation of the Kubo formula using the plane wave basis}
\label{s:kuboheuristics}
\noindent
 In the analytical derivation based on the Kubo formula \eref{kubo} presented in this section we will assume that $\sigma^\textrm{intra}$ contains all intraband contributions, in this case only the Drude conductivity $\Drude$, whereas $\sigma^\textrm{inter}$ contains all interband contributions. (As mentioned, support for this assumption will come from the Boltzmann approach.)     
We therefore choose $\eta$ such that we obtain $\sigma^\textrm{intra}\equiv \Drude$  and then derive  $\sigma^\textrm{inter}$ with the given value of $\eta$, hoping that we obtain a sensible result. This is one of the assumptions make the derivation a heuristic one.

What furthermore allows an analytical treatment but also contributes to the heuristic nature of the derivation is that we assume that we---rather than using the eigenstates of the full problem including impurities---can use the  the eigenstates  \eref{eigenstates} of the free Hamiltonian, that is the plane wave states,   which makes an analytic evaluation simple. At high densities, that is, at high  Fermi momenta, the use of plane wave states should be a good approximation as the weak impurities deform the eigenstates only slightly from the plane wave states. However, at small Fermi momenta, close to the Dirac point, one would expect the eigenstates to be substantially deformed by the impurities and we know of no solid argument for why the use of plane wave states would give sensible results also close to the Dirac point. Nonetheless, the outcome appears to be sensible. In any case, the derivation gives a nice introduction into the non-intuitive interband contributions to the conductivity.  

The calculation is done in the chiral basis. For the evaluation we use that 
\be{noname-09}
\begin{array}{ccll}
(s, \, s\pr)  &  \hspace{1cm} & \langle \vec{k}s  | \vkm_x | \vec{k\pr} s\pr \rangle    \langle \vec{k\pr }s\pr  | \vkm_x | \vec{k} s \rangle  &  \\
\hline
(+1,+1)    & & v_k^2 \, \delta_{\vec{k}\vec{k\pr}}    \delta_{\vec{k\pr }\vec{k}}  (\cos \theta)^2   & \textrm{intraband} \\
(-1,-1)    & & v_k^2  \, \delta_{\vec{k}\vec{k\pr}}    \delta_{\vec{k\pr }\vec{k}}  (-\cos \theta)^2    & \textrm{intraband} \\
(+1,-1)    & & v_k^2  \,  \delta_{\vec{k}\vec{k\pr}}    \delta_{\vec{k\pr }\vec{k}}  (-i\tfrac{\Nchiral}{\Ndisp} \sin \theta) (i \tfrac{\Nchiral}{\Ndisp}\sin \theta) & \textrm{interband}   \\
(-1,+1)    & & v_k^2 \,  \delta_{\vec{k}\vec{k\pr}}    \delta_{\vec{k\pr }\vec{k}}  (i\tfrac{\Nchiral}{\Ndisp} \sin \theta) (-i\tfrac{\Nchiral}{\Ndisp} \sin \theta) & \textrm{interband}    \, . \\
\end{array}
\ee
Thus, we can compactly summarize this as
\be{noname-10}
\sigma = -i \frac{\hbar e^2}{L^2} \sum_{ss\pr \vec{k}}  v^2  
\frac{\FD{\erg_k^s}-\FDf (\erg_{k }^{s\pr})}{\erg_k^s-\erg_{k}^{s\pr}} 
   \frac{
\left(\tfrac{\Nchiral}{\Ndisp}\right)^{1-ss\pr }(1+ss\pr \cos 2\theta )/2
   }{
   \erg_k^s - \erg_{k}^{s\pr} + i \eta 
   }\, .
\ee
For the intraband part $s\pr=s$  we find the zero $\erg_k^s-\erg_k^{s\pr}$ in the denominator. However, also the numerator contains a zero in $\FD{\erg_k^s}-\FDf (\erg_{k }^{s\pr})$. Since the arguments of the two Fermi-Dirac functions are 'close' to each other, we can Taylor expand the difference as
 \be{noname-02}
 \FD{\erg_k^s}-\FDf (\erg_{k }^{s\pr}) = 0 + (\erg_k^s-\erg_k^{s\pr}) \frac{\partial\FD{\erg_k^s}}{\partial \erg_k^s}     + \ldots
 \ee
 We see that the numerator contains the same zero as the denominator and we can cancel the two occurrences of  $\erg_k^s-\erg_k^{s\pr}$ to obtain the intralayer contribution 
 \be{noname-03}
\sigma_{\textrm{intra}} & =&  -i \frac{\hbar e^2}{L^2} \sum_{s\vec{k}}  v^2 
 \frac{\partial\FD{\erg_k^s}}{\partial \erg_k^s}   
   \frac{
 \cos^2 \theta 
   }{
   \erg_k^s - \erg_{k}^{s} + i \eta 
   } = 
   \frac{\hbar e^2}{\eta L^2} \sum_{s\vec{k}}  v^2  \cos^2 \theta 
\delta (\erg_\fermi -\erg_k^s) 
= \frac{\hbar e^2}{\eta L^2} \sum_{\vec{k}}  v^2  \cos^2 \theta 
\delta (\erg_\fermi  -\erg)  \longrightarrow \nonumber \\
 & \longrightarrow  & 
 \frac{\hbar e^2}{\eta } \int \frac{\der^2 k}{(2 \pi)^2}  v^2  \cos^2 \theta 
\delta (\erg_\fermi  -\erg) =
 \frac{\hbar e^2}{\eta } \int \frac{\der \theta}{2 \pi}  \cos^2 \theta \int \der \erg \DS (\erg) v^2
\delta (\erg_\fermi  -\erg) =
 \frac{\hbar e^2}{2\eta } \DS_\fermi  v_\fermi^2 
 % =  \frac{\hbar e^2}{2\eta }\frac{k_\fermi}{2 \pi \hbar v_\fermi } v_\fermi^2
  = 
 \frac{\hbar e^2}{2 h } \frac{\hbar }{\eta} k_\fermi v_\fermi\, ,
 \ee
where we used $\DS(\erg) =\frac{k}{2 \pi \hbar v} $ and where we have assumed zero temperature. 
We set $\eta = \hbar /\ttr$ in order for $\sigma^\textrm{intra}$ to reproduce the Drude conductivity $ \Drude$.

\begin{figure}
 \begin{center}
 \includegraphics[width=\textwidth]{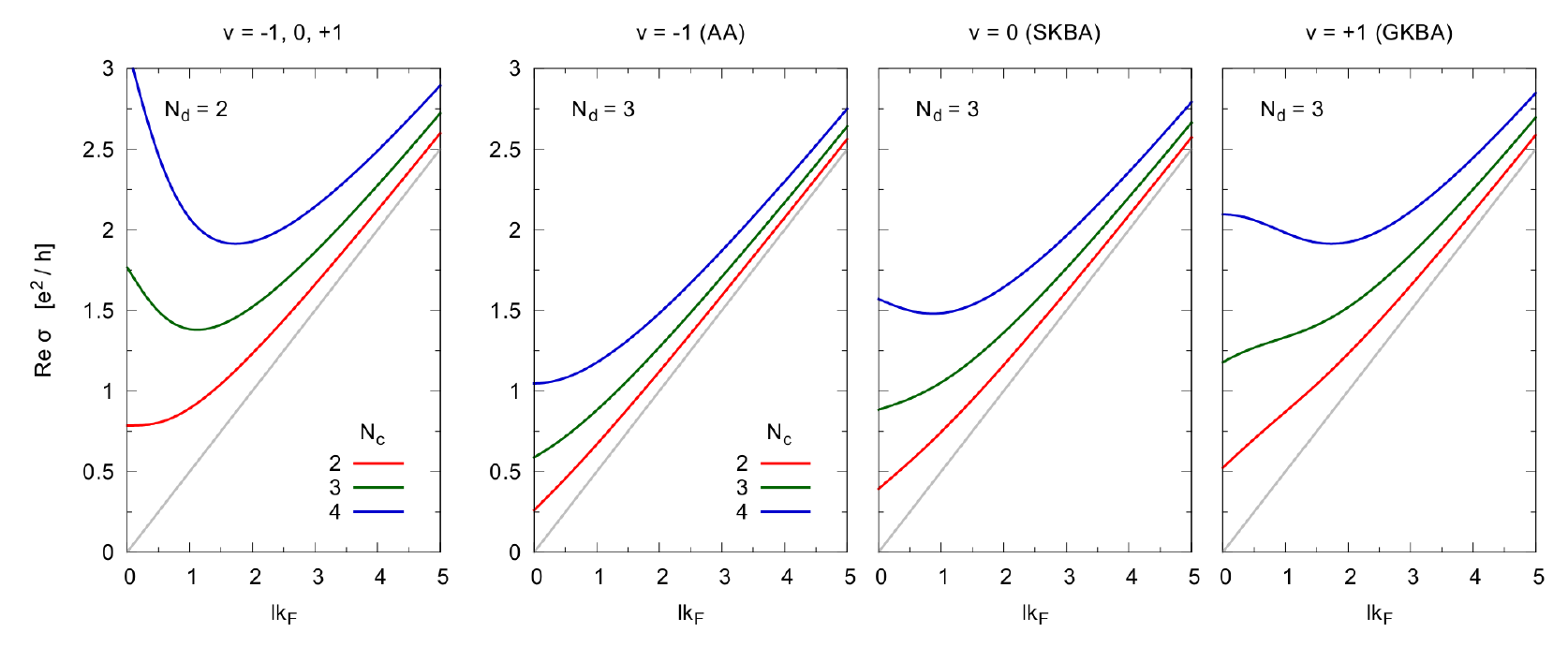}
 \caption{\label{fig1} The Boltzmann conductivity for different $\Ndisp$ and $\Nchiral$.
 The leftmost panel shows the case of $\Ndisp=2$ (quadratic dispersion), where all three Boltzmann approaches agree.
 The other three panels show the Boltzmann conductivity for $\Ndisp=3$ with different choices of Ansatz, that is, with different outcomes for the principal value terms in the collision integral.  
 The gray line corresponds to the Drude prediction.
 These results suggest that it is the increase chiral winding number $\Nchiral$that pushes up the conductivity 
 close to the Dirac point as on increases $N$, whereas the increase in density of states at the Dirac point that comes with increasing $\Ndisp$ instead decreases the conductivity. Note that this observation applies for all considered choices of Ansatz .}
\end{center}
\end{figure}

\begin{figure}
 \begin{center}
 \includegraphics[width=\textwidth]{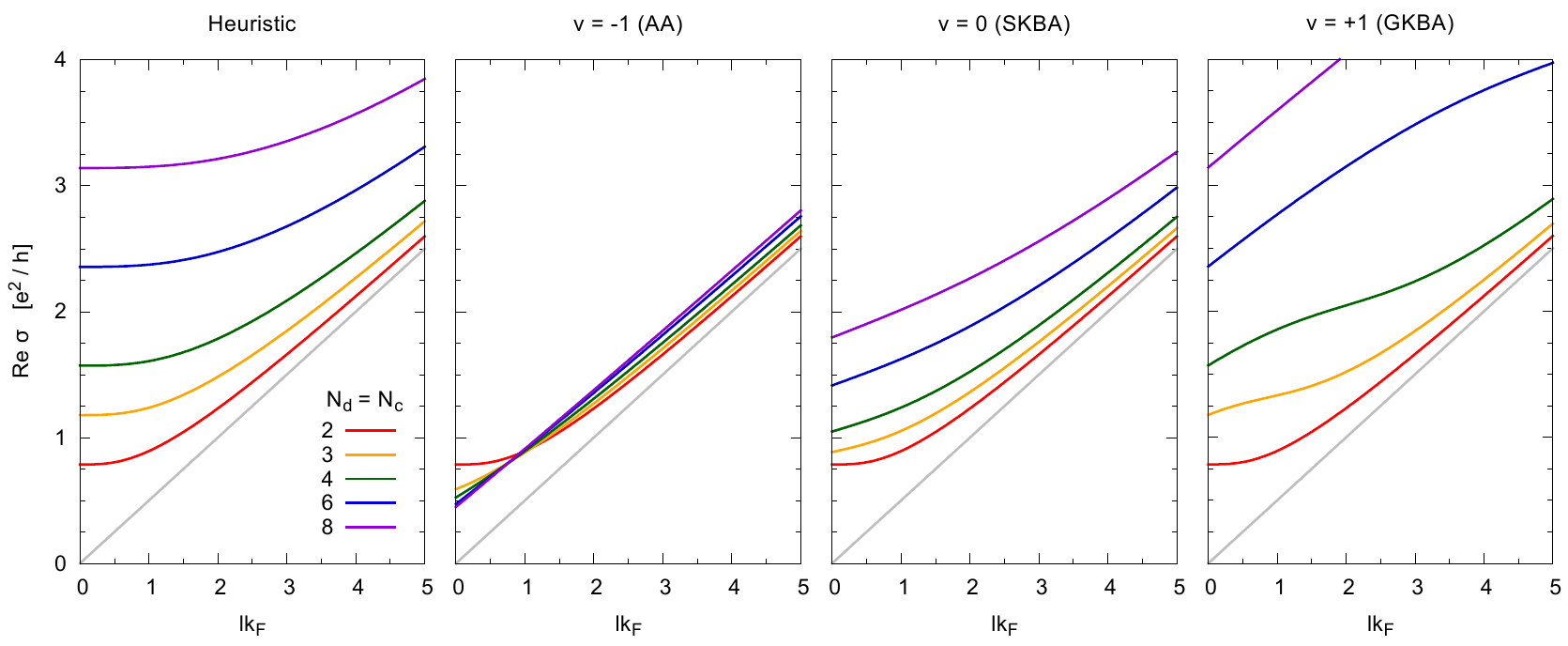}
 \caption{ \label{fig2} The analytical results for the total conductivity for $N(=\Ndisp=\Nchiral)$ equal to 2 (red), 3 (yellow), 4 (green),
 6 (blue), and 8 (magenta). The Drude prediction is shown by the gray line.
 The leftmost panel shows the results of the heuristic analytical treatment of the Kubo formula, and the subsequent three panels
 demonstrate the Boltzmann conductivities for a particular Ansatz.
 Note that the conductivity at the neutrality point  $\ell k_\fermi=0$ {\it decreases} with increasing $N$
 in the case of the Ansatz AA, but does not do so in the other considered cases. Note also that at $\ell k_\fermi =0$ the values with the Ansatz GKBA agree exactly with those with the heuristic treatment, but that for other $\ell k_\fermi$ there is no agreement for $N>2$.
 The further comparison with the numerically calculated conductivity will allow us to figure out,
 which Ansatz should be considered as the most proper one.}
\end{center}
\end{figure}

  We now turn to the interband contribution. We find  
 \be{noname-04}
\sigma_{\textrm{inter}} & =&    
-i \frac{\hbar e^2}{L^2} \frac{\Nchiral^2}{\Ndisp^2} \sum_{s \vec{k}}  v^2  
\frac{\FD{\erg_k^s}-\FDf (\erg_{k }^{-s})}{\erg_k^s-\erg_{k}^{-s}} 
   \frac{
\sin^2 \theta
   }{
   \erg_k^s - \erg_{k}^{-s} + i \eta 
   }
   = \nonumber \\
   & =& 
-i \frac{\hbar e^2}{L^2} \frac{\Nchiral^2}{\Ndisp^2}  \sum_{ \vec{k}}  v^2  \sin^2 \theta
\left [
\frac{
\Theta(\erg_\fermi-\erg ) -\Theta (\erg_\fermi +\erg)
}{
(2 \erg)
  (2\erg + i \eta) 
   }
  +
 \frac{
\Theta(\erg_\fermi+\erg ) -\Theta (\erg_\fermi -\erg)
}{
(-2 \erg)
  (-2\erg + i \eta) 
   }  
 \right]  
  = \nonumber \\
& =& 
-i \hbar e^2  \frac{\Nchiral^2}{\Ndisp^2}  \int \frac{\der \theta }{2\pi }   \sin^2 \theta \int \der \erg \DS(\erg) v^2
\left [\Theta(\erg_\fermi-\erg ) -\Theta (\erg_\fermi +\erg) \right ]
\left [ \frac{1}{
(2 \erg)
  (2\erg + i \eta) 
   }
  -
 \frac{
1
}{
(-2 \erg)
  (-2\erg + i \eta) 
   }  
 \right]  
  = \nonumber \\
& =& 
i \frac{ \hbar e^2}{2}  \frac{\Nchiral^2}{\Ndisp^2}  \int_{\erg_\fermi}^\infty \der \erg 
 \underbrace{
  \frac{\DS(\erg) v^2}{2\erg}
  }_{
  \Ndisp/4 \pi \hbar ^2 
  } 
\frac{-i 2 \eta }{ (2 \erg )^2 +\eta^2} 
= 
\frac{e^2} {2 h}  \frac{\Nchiral^2}{\Ndisp}  
 \int_{\erg_\fermi}^\infty \der \erg  
 \frac{ \eta }{ (2 \erg )^2 +\eta^2} 
= 
\frac{e^2}{4 h}  \frac{\Nchiral^2}{\Ndisp } \left [ \frac{\pi}{2} - \arctan \left ( \frac{2 \erg_\fermi}{\eta}  \right) \right] \, .
 \ee 
With the previously assumed $\eta = \hbar /\ttr$ we obtain  \be{kuboplanewavesigma}
\sigma_{\textrm{inter}}   =  
  \frac{e^2}{4 h}  \frac{\Nchiral^2}{\Ndisp}  \left [ \frac{\pi}{2} - \arctan \left ( \frac{2 \erg_\fermi \ttr }{\hbar}  \right) \right] 
  %= 
 % \frac{e^2}{4 h}  \frac{\Nchiral^2}{\Ndisp}  \left [ \frac{\pi}{2} - \arctan \left ( \frac{2 \ell k_\fermi }{N}  \right) \right]  
   \, .
  \ee 
 The total conductivity is therefore  
\be{kuboheuriresult}
\sigma = \sigma_{\textrm{intra}} +\sigma_{\textrm{inter}} = \frac{e^2}{2h} \ell k_\fermi  + \frac{e^2}{4 h}  \frac{\Nchiral^2}{\Ndisp}  \left [ \frac{\pi}{2} - \arctan \left ( \frac{2 \ell k_\fermi }{\Ndisp}  \right) \right]\ . 
\ee
The rewriting 
\be{noname-11}
\sigma_{\textrm{inter}} = \frac{e^2}{4 h}  \frac{\Nchiral^2}{\Ndisp} \arctan \left ( \frac{\Ndisp}{2 \ell k_\fermi }  \right)
\ee
with the arcus tangens function taylor expanded in powers $(\ell k_\fermi)^{-1}$ shows how an infinite series of terms of pseudospin originated quantum corrections diverging at $\ell k_\fermi =0 $ can sum up to something finite. In the Boltzmann regime  $\ell k_\fermi \gg 1$ the interband part vanishes and we are left with the Drude contribution. In the opposite limit $\ell k_\fermi =0 $ the Drude conductivity vanishes and the interband contribution saturates at the finite value  
\be{kuboplanewavemin}
\sigma  (\ell k_\fermi=0)
%_{\textrm{min}}   
=  
  \frac{e^2}{ h}
  \frac{ \pi}{8}  \frac{\Nchiral^2}{\Ndisp}  \, .
  \ee
  Note that the increasing the chiral winding number $\Nchiral$ pushes up the conductivity at the Dirac point, whereas increasing the power $\Ndisp$ of the dispersion pushes it down. 
For the multilayer Hamiltonian \eref{multilayer} ($\Nchiral=\Ndisp=N$) the total conductivity has a minimum at the Dirac point $\ell k_\fermi =0$, and the value of this minimum is proportional to the number of layers $N$. In particular, for the bilayer case $N=2$ this agrees with the result $\sigma_{\textrm{min}}   
=  
 \frac{e^2}{ h}
\frac{ \pi}{4} $
in ref.~\onlinecite{Trushin-Kailasvuori-Schliemann-MacDonald-2010} derived analytically with a spin-coherent Boltzmann type approach.  
In the case $\Nchiral > \Ndisp $ the qualitative behavior is different. A minimum occurs at some non-zero $\ell k_\fermi$ whereas close enough to $\ell k_\fermi =0$ the conductivity shoots up to saturate at a local maximum, with the value given by \eref{kuboplanewavemin}.  Examples are shown in Figs.~\ref{fig1} and \ref{fig2}.

\section{Detailed introduction to the background and motivation of the paper}
\label{s:background}
\noindent
This section complements the introduction with a more technical and detailed account of the background and motivation of the present study as for the methodological issues concerning the use of band-coherent generalizations of Boltzmann equations to study graphene-related systems.  

The Boltzmann equation is a kinetic equation that is a widely used starting point for deriving the transport coefficients and other response functions, also for quantum systems, where it is used as a semiclassical approximation. The appeal of a Boltzmann type of approach is that it makes contact with a classically intuitive picture, in contrast with the kind of numerical evaluation discussed later in the paper. Furthermore, many problems can for a "first feel" be suitably approximated to be amenable to an analytic solution. Both of these features offer great help for intuition. At the same time, the Boltzmann equation can be derived from a full many-body quantum theory, for example with the Kadanoff-Baym equations as the starting point. This means that it in principle should be possible to keep track of all the approximation and their range of validity, and it also means that there is a way to work out more refined and extended equations, when needed. This stands in contrast to a derivation such as the one 
in section~\ref{s:kuboheuristics}. However, whereas the Boltzmann approach is a very simple and instructive tool for simple problems such as deriving the Drude conductivity in an ordinary metal where spin is involved in a trivial way, it gets very complicated as soon as one considers more complicated systems such as graphene, where the pseudospin has a very non-trivial coupling with the momentum. Still, there is the possibility, as discussed in this paper, that a Boltzmann approach could make sense even for graphene-like systems and in some cases allow a fully analytical solution. This should be one clear motivation for investigating generalizations of Boltzmann equations in systems with non-trivial spin, in particular with some sort of spin-orbit coupling.

In this paper it is shown that the simple but due to their chirality yet highly non-trivial model Hamiltonians \eref{multilayer} with the generalization \eref{NdNc} provide an ideal testing ground for scrutinizing and screening the kind of band-coherent generalizations of the Boltzmann equation that have been used in refs.~\onlinecite{Auslender-Katsnelson-2007, Trushin-Schliemann-2007, Culcer-Winkler-2007b, Liu-Lei-Horing-2008,Culcer-Winkler-2008, Kailasvuori-Lueffe-2010, Trushin-Kailasvuori-Schliemann-MacDonald-2010} to discuss the conductivity in regime close to the Dirac point, that is,  at the low charge carrier filling where the chemical potential is close to the zero of energy given by the touching of  the two bands, the one with pseudo-spin up and the other with  pseudo-spin down along the momentum. In this regime it might be relevant to not only consider the energy eigenstates of the free Hamiltonian, which describe the population of respective pseudo-spin band, but also consider electrons in quantum 
coherent superpositions of these eigenstates.  (We will interchangeably use the terms spin-coherent and band-coherent to make it clear that we do not only consider populations of the eigenstates but  also quantum superpositions of these.) The mentioned band-coherent Boltzmann equation approaches treat the spatial degrees of freedom semi-classically as in the standard Boltzmann treatment, but keep the treatment of the the band index---that is, of the (pseudo)spin---fully quantum mechanical. Because of the band-coherent terms, which are elusive to classical intuition, the collision integral of these equation cannot sensibly be  given by an educated guess, but have to be derived from full quantum many-body theory.

We are interested in to what extent these kinds of generalized Boltzmann approaches are in principal applicable, considering that the usual textbook criterion $\ell k_\fermi \gg 1$ for a Boltzmann treatment breaks down close to the Dirac point. We are also interested in if this approach is in practice possible, considering that for Dirac systems the collision integrals in these approaches usually become horrendously complicated, in particular if one takes the often neglected principal value terms seriously. The latter turn the static and uniform Boltzmann equation---usually a differential equation problem---into a integro-differential problem. The mathematical challenges that come when solving these can be clearly seen  in  ref.~\onlinecite{Auslender-Katsnelson-2007}).

We show that the band-coherent Boltzmann equations considered in this paper  are analytically solvable for the Hamiltonians  \eref{multilayer} and \eref{NdNc} with the impurity potential given by point-like impurities,  and most remarkable the equations remain so with moderate additional complexity even when principal value terms are kept. The analytical form of these results makes the lack of equivalence of  the different Boltzmann approaches conspicuous and facilitate a neat matching against other approaches that are considered more reliable close to the Dirac point, in our  case a numerical evaluation of the finite-size Kubo formula. In the special case of bilayer graphene with point-like impurities, most of these approaches coincide. In this case the prediction of a band-coherent Boltzmann approach agrees well with the numerical evaluation, as shown in ref~\cite{Trushin-Kailasvuori-Schliemann-MacDonald-2010}.    

The source of this multiplicity among Boltzmann approaches comes from there being a wide variety of  derivations of Boltzmann equations from quantum many-body theory, with both different starting points and different approximations on the way. Even if the different approaches are sometimes motivated by different purposes, there are usually overlapping regimes where a comparison should be possible, like our   case of weak and dilute impurities (implying that it should be enough to consider the lowest Born approximation and only non-crossed diagrams). In the context of graphene and spin physics the issue of their equivalence, or rather lack of such, was addressed by one of us \cite{Kailasvuori-Lueffe-2010}.  It was shown that  although there turns out to be agreement in some standard electronic systems, there is no general equivalence. In particular,  for electronic  systems where coupling between momentum and some spin type of variable is strong---as is the case for the here studies Hamiltonians with the 
pseudo-spin to momentum coupled term being the dominant term---at least six feasible and yet different band-coherent generalizations of the Boltzmann equations can be derived, and, as  discussed in ref.~\onlinecite{Kailasvuori-Lueffe-2010}, they all result in different spin-originated quantum corrections to the conductivity. On the other hand, in ordinary textbook systems with all degrees of freedom treated semi-classically, in particular for essentially spinless electrons with quadratic dispersion like in the ordinary metal, these derivations all agree, which is probably one of the reasons for the issue of equivalence of Boltzmann approaches not having received much attention. 

This abundance of candidates for the Boltzmann equation comes above all from the non-equilibrium Green's functions derivations. Here the Kadanoff-Baym (or Keldysh) equation is the starting point. In the derivation of a Boltzmann equation one has to choose whether the real part of the self-energies should be related to relaxation (through the collision integral), as is the fate of the imaginary part, or if the real part instead goes into the renormalisation of the free drift of the semiclassical quasiparticle. The most important source of discrepancy, however, comes from the choice of \textit{Ansatz}.  The reason that one needs the discussed Ansatz is that one needs some approximation for expressing a two-time non-equilibrium Green's function $G(t_1,t_2)$ (other variables are here left implicit) in terms of of the one-time distribution function $f(t)$ in order to get a closed equation for latter. The necessity of the Ansatz is illustrated in appendix~\ref{s:whyansatz} in schematic explanation that has been 
abstracted and simplified from the kind of elements that occur in actual quantum many-body derivations of Boltzmann equations, see for example ref.~\onlinecite{Kailasvuori-Lueffe-2010} and references therein.   

The Green's function derivation can be contrasted with the more common and much more accessible quantum Liouville equation -based derivations of master equations (in our case the generalized Boltzmann equations) for the density matrix $\rho(t)\sim f(t)$.  Appendix~\ref{s:liouville} shows the origin of the Boltzmann collision integral in such a derivation. Here one is not confronted with the issues mentioned above. In the graphene related literature, almost all treatments have been of the Liouville type, see refs.  \onlinecite{Auslender-Katsnelson-2007, Trushin-Schliemann-2007, Culcer-Winkler-2007b, Culcer-Winkler-2008, Trushin-Kailasvuori-Schliemann-MacDonald-2010}, the to us known exceptions being refs. \onlinecite{Liu-Lei-Horing-2008, Kailasvuori-Lueffe-2010}.  In ref.~\onlinecite{Kailasvuori-Lueffe-2010}  we showed that none of the existing Ansatzes in the Green's function literature gave a collision integral in  the band-coherent Boltzmann equation that matched the collision integral we derived by a 
Liouville equation approach, and we invented yet another Ansatz to provide the missing link. With this we yet increased the number of Boltzmann equations that can be derived with a Green's function approach. 

One might think that a Liouville equation -based derivation is then the correct derivation as one does not seem to be confronted with these elements of choice. Our study does not support this view. Rather, we find that if any of the approaches can be ruled out, then it is the Liouville equation -based one, at least in its standard form. 

A central technical issue in the present paper is the role of principal value terms in the collision integral. They arise as byproduct in all derivations of Boltzmann equations mentioned above, also in the Liouville type of derivation. Principal value terms have been discussed in other contexts (under various names), for example the Boltzmann equations with spinless electrons subject to strong interactions, but have in the context of Dirac electrons and spin-orbit coupled electrons simply been left out, with the to us known  exceptions being refs.~\onlinecite{Auslender-Katsnelson-2007, Kailasvuori-Lueffe-2010}. The usual Boltzmann collision integrals have as typical ingredients a Dirac delta function $\delta( \erg_k^s-\erg_{k\pr}^{s\pr} )$ expressing the classically intuitive conservation of energy in a collision process. The quantum many-body derivations that lead to these expected collision terms will usually also result in similar terms, which instead of the delta functions involve principal values $\prip{
 (\erg_k^s-\erg_{k\pr}^{s\pr})^{-1} }$. (Appendix~\ref{s:liouville} shows explicitly where they come about). The latter are rather elusive to a semiclassical interpretation. They might be connected with the Zitterbewegung property of Dirac electrons and  spin-orbit coupled electrons. We believe that is motivated to address their presence and role seriously, in particular when the distribution describes quantum coherent superpositions of eigenstates.  

As mentioned above, the principal value terms make in most cases the Boltzmann equation horrendously complicated to solve. Maybe one reasons---on top of them being unintuitive---for them usually being neglected, most often without discussion. The special attractive feature of the models \eref{multilayer} and \eref{NdNc} together with point-like impurities is that there are non-vanishing principal value terms due to the band-coherence but an analytical solution of the different Boltzmann equations is nonetheless still possible to infinite order in $(\ell k_\fermi)^{-1}$ in spin-originated quantum corrections and sums up to a neat result that remains finite even at the Dirac point $\ell k_\fermi=0$. Remarkable here is that the presence of principal value terms only moderately but not qualitatively increases the complexity of the solution. Furthermore,  it turns out that the principal value terms modify the conductivity in a very non-trivial way. The studied setting is therefore a very interesting testing 
ground for screening of the different approaches and an addressing of the role and influence of principal value terms as well as the applicability of band-coherent generalizations of the Boltzmann equation.

\section{The band-coherent Boltzmann treatment}\label{s:boltzmann}
\noindent
In this section we are going to derive within the Boltzmann approach the pseudospin originated quantum corrections in the considered graphene type systems and in particular with principal value terms included. However, we have refrained from making the paper self-contained in this part, as this would make the paper far too ominous. For a technical understanding of the results familiarity is assumed with the derivation of Boltzmann equations in general as well as with ref.~\onlinecite{Kailasvuori-Lueffe-2010}, where technical details for the band-coherent case have been worked out in great detail and generality. This section functions as a guide for those readers who want to see how that work can be applied to the present context. Other reads might want to jump to the end of the section were the main results and conclusions have been gathered.   

In the Boltzmann approach the particles are described in terms of a distribution function $f(\vec{x},\vec{k},t)$ living in classical phase space. In the presence of spin this function obtains a spin index, usually without any semiclassical approximation, making $f$ a density matrix in the spin indices. The Boltzmann equation then reads
\be{lhs0}
i[H,f]+\partial_t f+\frac{1}{2}\{{\vkm}_i, \partial_{x_i} f\}+eE_i\partial_{k_i} f= \mathcal{J}[f]  \, ,  
\ee
with the bracket $\{\, ,\,\}$ standing for anti-commutation.
The presence of spin leads among other to the presence of the spin-precession term $i[H,f]$. However, spin-precession has also a classical analogue in the precession of angular momentum.  The truly quantum mechanical ingredients are found in the interband components of  $\{{\vkm}_i, \partial_{x_i} f\}$ and in the collision integrals $\JC[f]$, for example in Pauli blocking factors and in the collision rates.

 The Boltzmann equation can be derived from quantum many-body theory,  with the main issue being the derivation of the collision integral. One common derivation is to use the quantum Liouville equation as a starting point. A derivation of a collision integral can be found in appendix~\ref{s:liouville}. Another starting point is the Kadanoff-Baym equations, from which one can derive 
\be{kb1}
 [i\partial_t -H,\GL] ~=~ \SR\GL-\GL\SA +  \SL \GA -\GR \SL \, . 
\ee
After Wigner transformation followed by gradient expansion and  integration over the energy $\omega$ the left hand side turns into the left hand side of \eref{lhs0}, where by definition 
\be{defGf}
f (\vec{k},\vec{x},t)= \int \frac{\der \omega}{2\pi}  \GL (\omega,\vec{k},\vec{x},t)  \, , 
\ee  
An equivalent form of \eref{defGf} in a non-Wigner transformed coordinates is 
\be{noname-12}
f(x_1,x_2,t) = \GL(x_1,t_1,x_2,t_2)|_{t_1=t_2=t}\, , 
\ee
that is, the distribution function is related to the time-diagonal of the lesser Green's function. 
The right hand side of eq.~\eref{kb1}---involving self-energy terms $\Sigma [G]$ which are functionals of the non-equilibrium Green's function $G$---turns into something that we can interpret as a collision integral. 
To derive a collision integral in the lowest Born approximation, the dressed Green's functions $\GR$ and $\GA$ are replaced by their non-interacting analogues $\GoR$ and $\GoA$ and  in the self-energies the series in orders of the interaction is only kept to second order in the interaction and first order in the impurity density.  The retarder non-interacting Green's function is in our case given by 
\be{gor}
\GoR_\vec{k}(\omega)~ =~\sum_{s=\pm} \frac{\sproj_{\uv{\ben}s }}{\omega + i0^+ - \erg^s_\vec{k}} =
\sum_{s=\pm} \sproj_{\uv{\ben}s }
\left( 
\frac{\mathcal{P} }{\omega -\erg^s_\vec{k} } - i\pi \delta  (\omega -\erg^s_\vec{k}) \right)
\ee
with the projector
\be{noname-13}
\sproj_{\uv{\ben}s}~:=~ 
\frac{1}{2}(\idm+\vec{\sigma}\cdot s\uv{\ben}_\uv{k})\, .
\ee  
 The advanced Green's function is obtained by hermitian conjugation.

As mentioned above, the collision integral is the ingredient of the Boltzmann equation that is the most demanding to derive. One reason is that it is non-linear in the distribution functions. It is not surprisingly in the collision integral that we will find differences between the different derivations. We will let the  collision integral be divided up as $\JC=\JCd+\JCp$, with $\JCd[f]$ containing the terms that contain delta functions and $\JCp[f]$ containing the terms that contain principal value terms.  In the case of monolayer graphene (Hamiltonian \eref{multilayer} with $N=1$) both parts are sensitive to the approach, even for point-like impurities. (See sec. 6 in ref.~\onlinecite{Kailasvuori-Lueffe-2010}, in particular eqs. (65,66).) For the multilayer Hamiltonians with generalizations studied in the present paper, only the the principal value part will differ between the different approaches. 

Ref.~\onlinecite{Kailasvuori-Lueffe-2010} uses a nomenclature for the different approaches composed of a prefix and a suffix. The prefix \GO{} or \GT{} indicates whether the collision integral derives from all self-energy terms as suggested in equation \eref{kb1} or only from the imaginary part of the self-energies, namely from the right hand side of (the with \eref{kb1} equivalent expression) 
 \be{kb2} 
[i\partial_t-H-\Re \SR, \GL ]-[\SL, \Re \GR]~=~ i\{ \Im \SR, \GL\} -i\{\SL, \Im \GR\}\, .
\ee
 In the latter case the real part of the self-energies to the left are instead assumed to renormalize the free drift. (As ref.~\onlinecite{Kailasvuori-Lueffe-2010} was not able to carry out this renormalization in the spinful case, the statements and results for the approach $\GT$ have to be taken with more caution than those of $\GO$.) 
The suffix GKBA (Generalized Kadanoff-Baym Ansatz \cite{Lipavsky-spicka-Velicky-1986}), \AA{} ("Alternative Ansatz" or "Anti-ordered Ansatz" \cite{Kailasvuori-Lueffe-2010}) and \SKBA{} (Symmetrized Kadanoff-Baym Ansatz \cite{Kailasvuori-Lueffe-2010}) refers to which Ansatz has been used. In the Green's function derivations considered in this paper the issue of Ansatz is the central one.  

The Ansatz gives an explicit relation between the non-equilibrium Green's function $G$ and the distribution function $f$. By definition one has  $f(t)=\GL(t,t)$, or in the Wigner-transformed language equation \eref{defGf}. 
 This implicit relation is enough for the left hand side of \eref{kb1} to turn all Green's functions into distribution functions and this thanks to the left hand side being linear in Green's functions.  However, the right hand side is non-linear in Green's functions, wherefore the implicit relation such as \eref{defGf} is not enough, as explained in appendix~\ref{s:whyansatz}. Here one needs an explicit relation between $f$ and $\GL$ (and one needs explicit expressions for $\GoR$ and $\GoA$, here given by \eref{gor}). Such a relation between a semiclassical object and a fully quantum mechanical object has of course to be an approximation.  
 {\it The Ansatz} is the approximation that provides the explicit relation between $f$ and $\GL$, allowing one to  translate also in the other direction, that is, finding $\GL$ if one knows $f$. This is necessary for  distilling a closed equation for $f$---as is the Boltzmann equation---out of equation \eref{kb1}. (See again appendix~\ref{s:whyansatz} for a more detailed explanation.)
 
There are several possible candidates for the required approximation involved in the Ansatz.
The considered Ansatzes can for $\GL(\vec{x}_1,t_1,\vec{x}_2,t_2)$ in the lowest Born approximation be written as 
\be{ansatzes}
\begin{array}{lrcl} 
\textrm{GKBA}   &  \GL(t_1,t_2)  & = &  i \GoR (t_1,t_2) f(t_2) - if(t_1)\GoA(t_1,t_2) \\
\textrm{\AA} 			&  \GL(t_1,t_2)  & = &  i f(t_1) \GoR (t_1,t_2)  - i\GoA(t_1,t_2) f(t_2) \\
\textrm{\SKBA} 		&  \GL(t_1,t_2)  & = &  \frac{1}{2}  A^0(t_1,t_2) f(t_2) +\frac{1}{2}f(t_1) A^0(t_1,t_2) 
\end{array}
\ee
where $A =i (\GR-\GA)$ is the spectral function. All Ansatzes are consistent with $f(t) = \GL(t,t)$, but go beyond the definition by providing information also about the time-off-diagonal part of $\GL$. Note that the Ansatz \SKBA{} is the symmetrization of GKBA and \AA{}. In equation~\eref{ansatzes} the spacial two-point indices $(\vec{x}_1,\vec{x}_2)$ have been suppressed for convenience, and so has on the right hand side also the convolution product in these. (In the time indices there is no convolution product.) For the background to these Ansatzes, as well as for the approximations that follow upon the Wigner transformation and with the Born approximation, see ref.~\onlinecite{Kailasvuori-Lueffe-2010}. Here we just remention that the combination \GO \AA{} is what provides the missing link between the Green's function derived collision integral and the Liouville equation derived one  (as derived in appendix~\ref{s:liouville}), at least in the lowest Born approximation, which is what we consider. Let us 
also stress that the list of approaches need not be exhaustive, but provides only a set of natural candidates. 
 
The principal value part  $\JCp[f]$ of the collision integral can be written as $\JCp=\JCpx+\JCpy$ where $\JCpx[f]$ is the part that does not change sign when one compares the approach \GO GKBA to the approaches \GO\AA, whereas $\JCpy[f]$ is the part the  changes sign. The principal value terms $\JCp$ for the other approaches can be composed from these two parts and is given by the table  
\be{pripscheme}
\begin{array}{l l | l l}
&  & X & Y
 \\
 \hline
\textrm{\GO}  & \textrm{GKBA} & +  & + \\

\textrm{\GO}  & \textrm{\AA} &  + & - \\ 

\textrm{\GO}  & \textrm{\SKBA} & + & 0 \\

\textrm{\GT}  & \textrm{GKBA} &  0 & + \\

\textrm{\GT}  & \textrm{\AA} & 0  &  - \\

\textrm{\GT}  & \textrm{\SKBA} &  0 & 0  \, .
\end{array}
\ee
Using the matrix decomposition in eq.~\eref{fdecomp} the Boltzmann equation including the collision integral will in a linear approximation in the electric field decompose in the Fourier modes of the angular variables $\theta$ of the momentum $\vec{k}$. For the terms $\JCp[\fee]$, with $\fee$ being searched for linear-in-electric-field correction to the known equilibrium part $\feq$ one has according to ref.~\onlinecite{Kailasvuori-Lueffe-2010} (for notational convenience the superscript $(E)$ is suppressed; the prime superscript indicates that the component of $f$ is a function of $k\pr$ and not of $k$)\footnote{
In  ref.~\onlinecite{Kailasvuori-Lueffe-2010} as well as in the three first arXiv versionsof that paper the following formulae contained important mistakes. These mistakes turn out to have no impact on the results presented for the first quantum correction in monolayer graphene discussed paper. However, they have an important impact on the multilayer case studied in the present paper. The following equations are corrected equations, also to appear in corrected versions of the arXiv version of ref.~\onlinecite{Kailasvuori-Lueffe-2010} .
}
 \be{xXYF}
\begin{array}{rcl}
\JCpx_{\uv{\ben}1} & = & 0   
\\
\JCpy_{\uv{\ben}1} & = & +i\int_\vec{k\pr} \frac{\Wkk}{2\pi}
\pripm
\sin \Nchiral\DT \theta \sin\DT \theta f_{z1}\pr ,
\\   
 \JCpx_{\uv{\cen}1} &=&-\int_\vec{k\pr}  \frac{\Wkk}{2\pi}
\left( 
- \pripm \cos \Nchiral \Delta \theta f_{z1}
+
\pripp \cos \Delta \theta f_{z1}\pr
\right)  
 \\
 \JCpy_{\uv{\cen}1} &=&-\int_\vec{k\pr}  \frac{\Wkk}{2\pi}
\left( 
- \pripp  f_{z1}
+
\pripm \cos \Nchiral \Delta \theta \cos \Delta \theta f_{z1}\pr
\right)  
 \\
 \JCpx_{z1} &=& -\int_\vec{k\pr}  \frac{\Wkk}{2\pi}
\left \{
\pripm 
\cos \Nchiral\Delta \theta  f_{\uv{\cen}1}
 - 
\pripp
(
i \sin \Nchiral\Delta \theta \sin \Delta \theta  f_{\uv{\ben}1}\pr + \cos \Nchiral\Delta \theta \cos  \Delta \theta  f_{\uv{\cen}1} \pr
) 
\right\}
 \\
  \JCpy_{z1} &=& -\int_\vec{k\pr} \frac{\Wkk}{2\pi}
\left(
 \pripp f_{\uv{\cen}1} - \pripm \cos \theta \Delta f_{\uv{\cen}1}\pr 
\right)  
\end{array}
\ee  
with 
\be{noname-14}
\prippm=\frac{\mathcal{P}}{\ben+\ben\pr} \pm \frac{\mathcal{P}}{\ben - \ben\pr} \, .
\ee 
In the case of monolayer graphene, both $\JCpx$ and $\JCpy$ are typically non-zero, and the six candidates in \eref{pripscheme} lead  to six results---all different---for the leading quantum correction.  
In contrast, for the case of $\Nchiral \geq 2$ (including the multilayer Hamiltonians \eref{multilayer} with $N\geq 2$) combined with point-like impurities ($\Wkk=2\pi u_0^2=:W_0$) most of the terms in \eref{xXYF} vanish trivially, namely all the terms that contain trigonometric functions. This leaves as only nonzero terms
\be{JCpmulti}
\begin{array}{rclcl}
 \JCpy_{\uv{\cen}1} &=&+   f_{z1}\int_\vec{k\pr}  \frac{\Wkk}{2\pi}  \pripp 
 & =:  & +f_{z1} \Lambda_k
\nonumber 
\\
  \JCpy_{z1} &=& - f_{\uv{\cen}1} \int_\vec{k\pr} \frac{\Wkk}{2\pi}  \pripp 
  & =:  & - f_{\uv{\cen}1} \Lambda_k \, .
\end{array}
\ee
The vanishing of all terms $\JCpx$ implies  that we cannot distinguish between the approaches $\GO$ and $\GT$. We henceforth assume that we deal with $\GO$ and suppress this prefix. More interestingly,  the fact that some terms $\JCpy$ survive means that we can distinguish between the different Ansatzes GKBA, \AA{} and \SKBA. 
Furthermore,  the integral in $\Lambda_k$ can be worked out analytically as (see appendix~\ref{s:boltzcalc})
\be{noname-15}
\Lambda_k &=& \frac{2 W_0}{(2\pi)^2 \ben} \int_0^\infty k\pr\der k\pr \, \frac{\mathcal{P}}{1-\left (\frac{\ben\pr}{\ben}\right)^2}
=
 \frac{\cot \frac{\pi}{\Ndisp}}{\ttr} \left(\frac{k}{k_\fermi}\right)^{2-\Ndisp}
 \, .
\ee
Note that the derivation requires $\Ndisp >1$.   

Since the surviving principal value terms only involve $f_k$ but not $f_{k\pr}$ we can plug out the distribution function out of the integrals, which allows us to turn the integro-differential equations into differential equations, as is typically the case for static and uniform Boltzmann equations.  Thanks to this we essentially have the same type of Boltzmann equation as when the principal value terms were neglected. Section 6  in ref.~\onlinecite{Kailasvuori-Lueffe-2010} treated the latter problem in a general setting and the results can be easily translated to the present case. Equation (69) in ref.~\onlinecite{Kailasvuori-Lueffe-2010} is in the present case replaced by the following equation 
\be{matrixbe}
\frac{E_x-i E_y}{2}
\mtrx{c}{\partial_k \feq_0 \\  
\partial_k \feq_\uv{\ben} \\
i\tfrac{\Nchiral }{k} \feq_\uv{\ben} \\ 0 } 
=-
\mtrx{cccc}{
\IC  & 0 & 0 & 0 \\
0 & \IC & 0 & 0 \\
0 & 0  & \IC & 2\ben-\Ansatzindex \Lambda \\
0 & 0 & \Ansatzindex\Lambda-2b & \IC}
\mtrx{c}{\fee_{01} \\ \fee_{\uv{\ben}1} \\ \fee_{\uv{\cen}1} \\ \fee_{z1} }
\, .
 \ee
Most importantly one has the simple shift $2\ben \longrightarrow 2\ben -\Ansatzindex \Lambda$ in the strength of the precession term due to the principal value terms, with the sign depending on Ansatz according to
\be{noname-16}
\begin{array}{lrcl}
\textrm{GKBA} & \Ansatzindex & = & +1 \nonumber \\
\textrm{\AA} & \Ansatzindex & = & -1 \nonumber \\
\textrm{\SKBA} & \Ansatzindex & = & 0 \, .
\end{array}
\ee
Furthermore, $N\longrightarrow \Nchiral$ in the right hand side. The chiral origin of that $N$ can be found above equation (65) in ref.~\onlinecite{Kailasvuori-Lueffe-2010}. Other changes are that $\IC^+$, $\IC^\lambda$ and $\IC^\kappa$ coalesce to $\IC^+$, in the present paper denoted $\IC$, and that $\IC^s=0$ in the present setup.
The result in equation (75) in ref.~\onlinecite{Kailasvuori-Lueffe-2010} becomes in the present case 
 \be{sigma0sigma}
\begin{array}{rclcl}
\sigma^I & = & 
 \Drude  \,
\\
\sigma^{II} & =& 
%\frac{e^2}{2h} 
\frac{1}{4 \pi} \Nchiral^2 \int_{k_\fermi}^\infty  \der k \, \frac{\ben }{k} \frac{\IC}{(2 \ben-\Ansatzindex\Lambda)^2 +\IC^2} 
  \, .
  \end{array}
\ee 

The result that there are no quantum corrections to $\sigma^I$ gives an explanation from the perspective of the Boltzmann approach to why one of the central assumptions in section~\ref{s:kuboheuristics} made sense. There it was assumed that $\sigma^\textrm{inter}$ contains all the quantum corrections due to interband effects, whereas $\sigma^\textrm{intra}$ is identical to the Drude conductivity $\Drude$, so contains no corrections. The contribution  $\sigma^\textrm{inter}$ relies on the term $\vkm^\textrm{inter}$. However, according to definition of current in terms of the Boltzmann distribution fucntion in equation \eref{current}, the velocity component $\vkm^\textrm{inter}$  is what the distribution function  component $f_\uv{\cen}$ combines with. In the present case this component authors singel-handedly the correction $\sigma^{II}$. To see this, look at eq. (72) and eq. (75) in ref.~\onlinecite{Kailasvuori-Lueffe-2010} and remember that $\IC^s=0$ in our case.  Note that for the monolayer problem there 
is no such clean separation. There also $f_{\uv{\ben}}$, which contributes to the current through $\vkm^\textrm{intra}$, contains quantum corrections, wherefore we expect that for the monolayer case the crucial assumption in section~\ref{s:kuboheuristics} not to be valid. 

Even in the presence of principal value terms the correction term $\sigma^{II} $ can be rendered on a neat analytical form. With $e$ and $\hbar$ reintroduced, we arrive at the main result (see appendix~\ref{s:boltzcalc})
\be{reswprip}
\sigma^{I} & =& \Drude =  \frac{e^2}{ h} \frac{\ell k_\fermi }{2}=  \frac{e^2}{h} \frac{\Ndisp \ben_\fermi \ttr}{2\hbar}
\nonumber \\
\sigma^{II} &= & 
\frac{e^2}{h}\frac{\Nchiral^2}{8 (\Ndisp-1) } \left (\frac{\pi}{2}-\arctan\left(  \frac{2\ell k_\fermi}{ \Ndisp} - \Ansatzindex\cot \frac{\pi}{\Ndisp}\right)\right) 
\, .
\ee 
For $\Ansatzindex=0$ and $\Nchiral=\Ndisp$ this is the result in  equation (78) in ref.~\onlinecite{Kailasvuori-Lueffe-2010}. 

The conductivity at the Dirac point is found by setting $\ell k_\fermi=0$. The conductivity is then only composed of $\sigma^{II}$, since  $\sigma^{I}=0$. Using that 
\be{noname-17}
\arctan \left (\cot \frac{\pi}{N} \right) = \frac{N-2}{2 N} \pi 
\ee 
one has 
\be{noname-18}
\sigma(\ell k_\fermi =0) = \frac{e^2}{h}\frac{\pi}{2}\frac{\Nchiral^2(\Ndisp+\Ansatzindex (\Ndisp-2))}{8 (\Ndisp-1)\Ndisp } 
=
 \frac{e^2}{h}\frac{\pi}{8} \left\{
\begin{array}{ll}
\frac{\Nchiral^2}{ \Ndisp}  & \textrm{GKBA} \\
\frac{\Nchiral^2}{2 (\Ndisp-1)}  & \textrm{\SKBA} \\
\frac{\Nchiral^2}{ \Ndisp (\Ndisp-1)}  & \textrm{\AA} \, .
\end{array}
\right.
\ee

Here we gather the main points of this section
\begin{itemize}
\item
That all terms $\JCpx$ in equation \eref{xXYF} are zero means according to the scheme \eref{pripscheme} that the multilayer problem with point-like impurities does not allow to distinguish between all the approaches referred to in table \eref{pripscheme} . However, interestingly it is $\JCpy$ that survives, which according to \eref{pripscheme} allows us to distinguish between the three Ansatzes GKBA, \AA{} and \SKBA. 
\item
In contrast to the monolayer problem, the surviving principal terms depend only on $f_k$, but not on $f_{k\pr}$. One can therefore  plug out the distribution function out of the principal value integral, leaving a momentum dependent integral $\Lambda_k$ that is independent of the distribution function.  We are therefore luckily not forced to solve the kind of integro-differential equation considered by Auslender \etal \cite{Auslender-Katsnelson-2007}, but only the differential equations of the level of complexity that we have without principal terms. This is an appealing feature of the multilayer problem (and the general Hamiltonian \eref{NdNc} with $\Nchiral, \Ndisp\geq 2$) combined with point-like impurities. (Note that if the impurities are not point-like, then the terms with trigonometric functions in \eref{xXYF} normally do not vanish and we are stuck with the full integro-differential equations.)  
\item
An interesting special case is the one with quadratic dispersion $\Ndisp = 2$ but general chirality $\Nchiral \geq 2$. Then $\Lambda=0$ and the principal value terms vanish all together, with all approaches agreeing, see figure~\ref{fig1}.
 This is the fortunate circumstance that allowed us to neglect principal value terms and the question of different approaches in the case of  bilayer graphene with point-like impurities studied in ref.~\onlinecite{Trushin-Kailasvuori-Schliemann-MacDonald-2010}. It also results in density of states independent of filling, which we find is a big advantage for the numerical analysis of the Kubo formula as performed in ref.~\onlinecite{Trushin-Kailasvuori-Schliemann-MacDonald-2010} and in the next section of the present paper.  One important reason for introducing the artificial Hamiltonian \eref{NdNc} is to be able to study the role of increasing chiral winding numbers in such a special setting. 
\item
The Boltzmann derivation supports the crucial but loose assumption made in section~\ref{s:kuboheuristics} that all the quantum corrections are related to the interband velocity component $\vkm^\textrm{inter}$, contributing to $\sigma^\textrm{inter}$, whereas $\sigma^\textrm{intra}$, paired with the velocity component $\vkm^\textrm{intra}$,  comes with no corrections. See the discussion under equation~\eref{sigma0sigma}.
\item 
It is interesting to note that at $\ell k_\fermi =0$ the conductivity derived with the Ansatz GKBA in the Boltzmann approach matches the conductivity \eref{kuboheuriresult} derived in section.~\ref{s:kuboheuristics} with the heuristic analytic treatment of the Kubo formula. However, for $\ell k_\fermi >0$ the two results no longer agree except in the special case of quadratic dispersion $\Ndisp=2$, see figures~\ref{fig1} and \ref{fig2}. Thus, in general the heuristic derivation finds no equivalent among the discussed Boltzmann derivations.   
\item
As mentioned already in  section.~\ref{s:kuboheuristics} the conductivity $\sigma(\ell k_\fermi =0)$ is also the conductivity minimum when $\Nchiral \leq \Ndisp$. However, it is not necessarily so when $\Nchiral > \Ndisp$, in which case the conductivity minimum can occur at some strictly positive $\ell k_\fermi$ whereas $\ell k_\fermi$ corresponds to a local maximum in the conductivity as a function of $\ell k_\fermi$, see figure~\ref{fig1}. 
\item
It is interesting that the result derived with the Ansatz \AA, which would also be the result derived with a Liouville equation derivation, predicts a conductivity minimum that {\it decreases} with the number of layers $N(=\Nchiral=\Ndisp)$ of rhombohedrally stacked graphene, reaching $\sigma_\textrm{min}=\tfrac{e^2}{h}\tfrac{\pi}{8}$ (half of the bilayer result) in the limit $N\rightarrow \infty$. The other Ansatzes and the heuristic Kubo derivation predicts that the conductivity minimum increases with the number of layers, see figure~\ref{fig2}.   
\end{itemize}   
As one further point we mention that based on particle-hole symmetry we would expect the curve to be mirror symmetric if we were to plot $\sigma$ as a function of charge carrier density $n$ (which is negative for holes). However, since for positive $n$ we have $n \propto \ell k_\fermi$ in the studied setups, this would mean that if the conductivity at $\sigma(\ell k_\fermi)$ is not horizontal at $\ell k_\fermi=0$, then $\sigma(n)$ has a kink there where it passes $n=0$. However, such a kink does not seem very feasible and the numerical analysis does not suggest any kinks. As for this issue the analytical Boltzmann treatments discussed by us should not be expected to give a valid description at $\ell k_\fermi =0$. It is interesting to note that  the heuristically derived result based on the Kubo formula produces no such kink for $\Nchiral=\Ndisp$.

\section{Numerical study of the Kubo formula}\label{s:num} 

  \subsection{General issues} \label{s:generalissues}
\noindent
In this section we present the results obtained by a numerical evaluation of the Kubo formula \eref{kubo}. The Kubo formula can be applied to a finite system connected to leads by choosing $\eta= \gT \delta \erg$. The finite $\eta$ expresses the broadening of levels in a finite system given by a possibility of the particle to escape from the system and is related to the average energy spacing $\delta \erg = (\DS L^2)^{-1}$ at the Fermi energy. Here, $\DS$ is the density of states at the fermi surface. In the non-interacting case it is given by 
\be{}
\DS =  \frac{1}{2\pi \hbar^2  \cH \Ndisp  } p_\fermi^{2-\Ndisp}\, . 
\ee
$\gT$ is the dimensionless Thouless conductivity. In principle $\gT (\erg_\fermi)$ can be estimated. 
However, in the case that system is dirty enough so that the collision time $\tau_0$, which in our case equals $\ttr$, is much smaller than the Thouless escape time $\tTh = \hbar / (\gT \delta \erg)$, it turns out that, whereas the conductivity goes to zero for $\gT$ too small or too large,  it is rather insensitive to the precise value of $\gT$ in an broad intermediate range of the latter. 
As we are interested in bulk conductivity and do not want to probe the finiteness of the system and the presence of leads, it is exactly the conductivity in this $\gT$-insensitive plateau that we are after.  As in ref.~\cite{Trushin-Kailasvuori-Schliemann-MacDonald-2010, TrushinEPL2012} we typically set $\gT$ to be about 10.
%, which is sufficiently bigger than $\gT=1$ for the states to have enough overlap in energy, but still small enough for us not to violate the condition for the system being dirty enough
The condition for a dirty system can also be written as  
\be{noname-37}
1 \ll \frac{\tTh}{\ttr}  = \frac{\hbar \DS L^2}{\gT} \cdot \frac{2\pi }{\hbar } \frac{\DS n_i u^2}{2} = \frac{ \pi  \DS^2 N_i u^2}{\gT}  
\ee 
\ie
\be{e:condition1}
u \gg \sqrt{ \frac{\gT}{\pi N_i}  } \frac{1}{\DS}\, .   
\ee

We see that to assure \eqref{e:condition1}, the impurities should be strong enough, or the number of them has to be big enough. On the other hand, since we are focused on comparing with Boltzmann type equations  that were derived in the lowest Born approximation, it is important that the interaction strength is chosen small enough so that higher Born corrections can be expected to be negligible. Since higher Born corrections typically come with factors of 
\be{noname-20}
\sim \frac{U_{\vec{k}\vec{k\pr}}}{\erg_{\vec{k}}-\erg_{\vec{k\pr}}} 
\ee
the leading contributions will for point-like impurities $U_{\vec{k}\vec{k\pr}}=u_0/L^2$ come from the smallest denominators $\erg_{\vec{k}}-\erg_{\vec{k\pr}}$. An estimate of the average minimal size of this difference is given by average energy spacing $(\DS L^2)^{-1}$. We have therefore the second condition  
\be{e:condition2}
U_{\vec{k}\vec{k\pr}} = \frac{u}{L^2} \ll \delta \erg = \frac{1}{\DS L^2} \hspace{1cm} \Longrightarrow \hspace{1cm}  u \ll \frac{1}{\DS} 
\ee
Putting \eqref{e:condition1} and \eqref{e:condition2} together we obtain the following double condition on the parameters 
\be{e:condition3}
\sqrt{ \frac{\gT}{\pi N_i}  } \ll u D \ll 1.
\ee
We see that we need to have $\sqrt{\gT /\pi N_i} \ll 1$, which is rather easy to satisfy. The challenge is to keep $u D$ within the two inequalities in the condition \eref{e:condition3}. 
For ordinary point-like impurities the meeting of this condition will turn out to require an individual adjustment for each different $\Ndisp$. For the sake of comparability of different 
$N$ we have, for reason soon to be explained, ad hoc introduced an artificial modification of the point-like potential, with the strength being dependent of  $k_\fermi$.
The two point-like potentials $u$ which we will study numerically are 
\be{noname-24}
u = \left \{
\begin{array}{ll}
\frac{\Ndisp \cH \cp^\Ndisp L^2}{2\pi } \cD & \textrm{ordinary potential}\\
\frac{\cD}{\DS} & \textrm{artificial potential} \, .
\end{array}
\right. 
\ee
Here $\cD$ is a dimensionless constant and $\cp = 2\pi \hbar  /L$. For $\Ndisp = 2 $ the two potentials coincide.

 For a general dispersion the density of states $\DS$ depends on $k_\fermi$.  In the case of bilayers ($N=2$) studied in ref.~\cite{Trushin-Kailasvuori-Schliemann-MacDonald-2010} the situation was particularly simple, because for a quadratic dispersion $\erg \propto k^2$ the density of states  $\DS$  is a constant.  A set of parameters satisfying the condition \eqref{e:condition3} for some $k_\fermi$ satisfied the condition then for all $k_\fermi$. In contrast, in the multilayer case  with the dispersion $|\erg| \propto k^N$, the density of states $\DS$ depends on $k_\fermi$. For the ordinary potential with a constant strength this implies, that for low or high enough $k_\fermi$ left or right part of the condition will be increasingly violated.  At the low enough $k_\fermi$ the assumption of no higher Born corrections is violated, at the high enough $k_\fermi $ the assumption $\ttr \ll \tTh$ of the system being dirty is violated. The strength of the ordinary potential has therefore to be adjusted with a great care with respect to the power of dispersion as well to interval of $k_\fermi$ that we wish to study. We have introduced the  artificial potential to carry over the fortunate circumstances for bilayers to all $N$: with the artificial potential the condition \eqref{e:condition3} is satisfied for all $k_\fermi$ equally and this for all $N$ such that the other parameters, for example $\cD$, do not need to be adjusted as we increase $N$. 

We define the dimensionless momentum $q = (\hbar / \cp) k$ with its special value $N_e=\pi q_\fermi^2$. The impurity potential then reads
\be{e:impurity-potential}
U_k = \frac{u}{L^2} =  \frac{\Ndisp \cH \cp^\Ndisp}{2\pi } \left \{ 
\begin{array} {ll}
\cD  & \textrm{ordinary potential} \\
\cD q_\fermi^{\Ndisp-2}  & \textrm{artificial potential}
\end{array}  
\right..
\ee
The functional dependence of $\ell k_\fermi$ on the parameters for the two types of potential is
\be{ellkFfromq}
\ell k_\fermi = \left \{  
\begin{array}{ll}
\frac{2}{\cD^2 N_i } q_\fermi^{2\Ndisp-2}   & \textrm{ordinary potential}
\\
\frac{2}{\cD^2 N_i } q_\fermi^2  =  \frac{2}{\pi \cD^2} \frac{N_e}{N_i} =  \frac{2}{\pi \cD^2} \frac{n}{n_i}
 & \textrm{artificial potential}
\end{array}
\right..
\ee
Note that with the artificial potential we recover the bilayer situation that $\ell k_\fermi \propto n/n_i$, which is not the case for ordinary point-like impurities in the multilayer case.

Independent on the setup of the potential strength there is an additional upper constraint on $\ell k_\fermi$ coming with the momentum cutoff introduced by the necessarily finite matrices in the numerical treatment. The above defined dimensionless momentum $q$ is for a finite system integer valued and label the momentum states $p_i = \frac{2 \pi \hbar }{L} q_i$. For matrices of rank $2 q_\lambda^2$ (2 coming from the spin) we have to keep $\ell k_\fermi < \ell_\lambda k_\lambda$, where the cutoff is given by
\be{noname-27}
\ell_\lambda k_\lambda = \left \{  
\begin{array}{ll}
\frac{2}{\cD^2 N_i } q_\lambda^{2\Ndisp-2}   & \textrm{ordinary potential}
\\
\frac{2}{\cD^2 N_i } q_\lambda^2   & \textrm{artificial potential}
\end{array}
\right..
\ee

\subsection{Numerical results}

\begin{figure}
 \begin{center}
 \includegraphics[width=\textwidth]{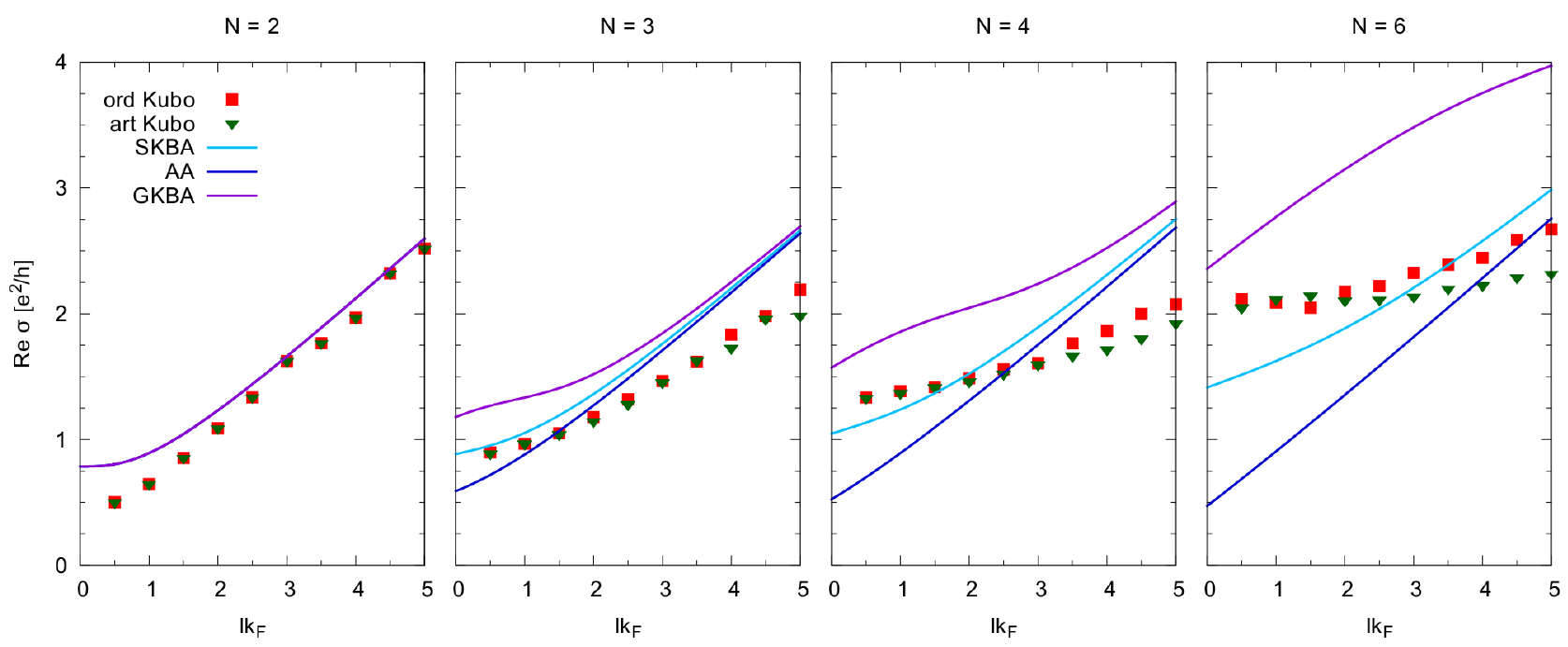}
 \caption{\label{fig3} The numerical finite-size Kubo conductivity computed for the chiral particles for a given $N(=\Ndisp=\Nchiral)$ in the presence of  ordinary point-like impurities (red squares) and an artificial scattering potential (green triangles), respectively.  The solid curves show the results for the different analytical Boltzmann approaches, distinguished by the Ansatz used for the derivation of the collision integral. The analytical curve in the middle corresponds to the Ansatz SKBA in which principle value terms are absent in the collision integral. Of the compared Boltzmann approaches this one appears come closest to  the numerical results in the range of low momenta.  The parameters used for the conductivity computations in the presence of the artificial potential are $q_{\lambda} = 14$, $\cD = 0.4$, $g_{\rm T} = 12$ at all $\Ndisp$. With the artificial potential the obeying of  condition \eqref{e:condition3} is independent of the value of  $k_F$ as well as of $N$. 
  In the case of the ordinary point-like impurities, the parameters must be chosen different for each $\Ndisp$ to satisfy condition \eqref{e:condition3}. We have used $\Ndisp =2$:
  $q_{\lambda} = 14$, $\cD = 0.4$, $g_{\rm T} = 12$; $\Ndisp = 3$: $q_{\lambda} = 12$, $\cD = 2.4$, $g_{\rm T} = 12$; $\Ndisp=4$:
  $q_{\lambda} = 12$, $\cD = 10$, $g_{\rm T} = 12$; $\Ndisp =6$: $q_{\lambda} = 10$, $\cD = 200$, $g_{\rm T} = 12$.}
\end{center}

\end{figure}

\begin{figure}
 \begin{center}
 \includegraphics[width=\textwidth]{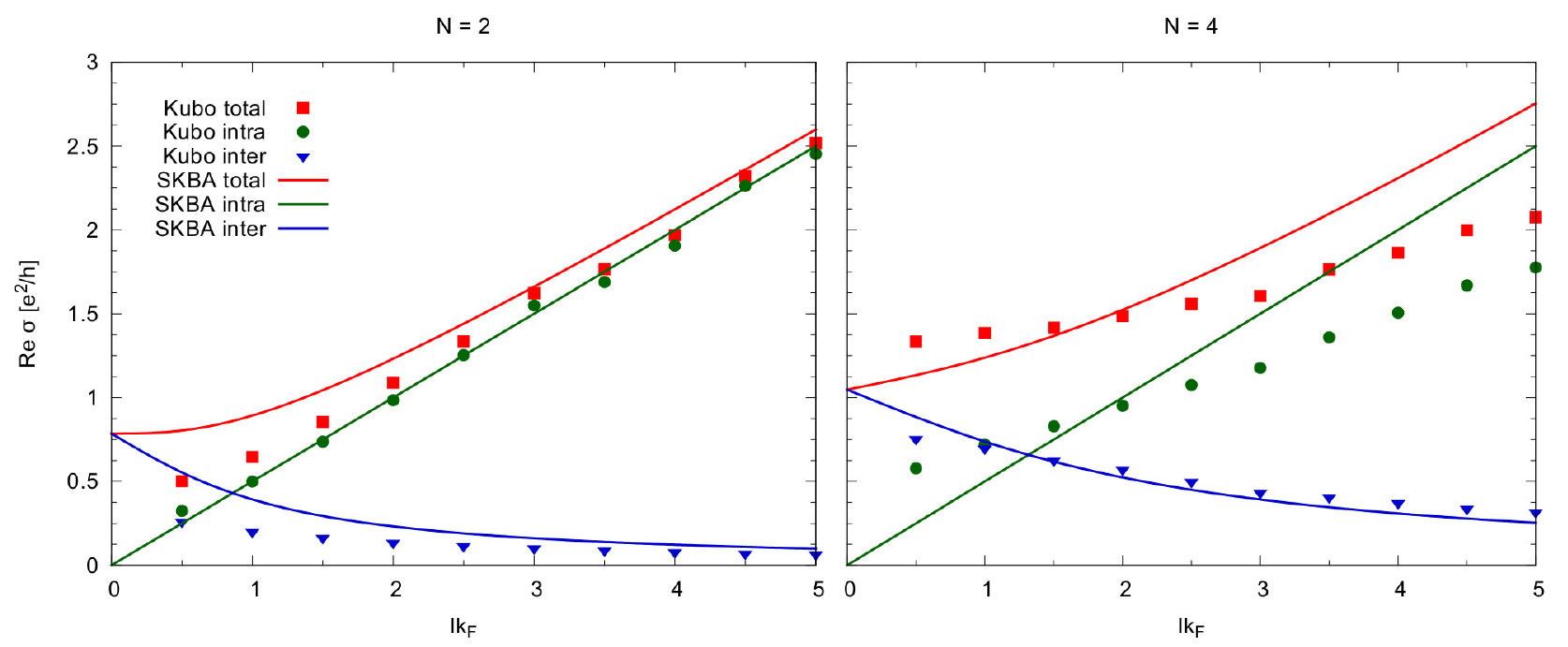}
 \caption{\label{fig4} The numerical Kubo conductivity (red squares) computed for the chiral particles with $N=2$ and $N=4$, respectively,   in the presence of ordinary point-like impurities.  Also shown is the intraband component $\sigma^\textrm{intra}$ (green circles) and the interband component $\sigma^\textrm{inter}$ (blue triangles)  The qualitative behavior is matched by the analytical formulae derived within the Boltzmann approach with the Ansatz SKBA. The parameters used for the conductivity computations are the same as in the previous figure.}
\end{center}

\end{figure}

\noindent
We will now discuss our numerical results presented in Figures~\ref{fig3} and \ref{fig4}. The plots show the conductivity in units of $e^2/h$ as a function of the dimensionless variable $\ell k_\fermi$. We consider two cases of disorder: ordinary point-like impurities as well as the artificial point-like impurities, as introduced in Sec.~\ref{s:generalissues}. 
Figure~\ref{fig3} shows the total conductivity for $N = \Ndisp = \Nchiral$ equal to 2, 3, 4 and 6, with the corresponding analytical dependence given by the solid curves.
The numerical results are consistent with the analytical prediction that the conductivity does not fall down to zero at  $\ell k_\fermi=0$, i. e. the main qualitative feature -- the finite conductivity minimum at the neutrality point -- is captured. Figure~\ref{fig4} shows the intra- and interband contributions $\sigma^\textrm{intra}$ and $\sigma^\textrm{inter}$. The intraband contribution $\sigma^\textrm{intra}(\ell k_\fermi)$ is  monotonically increasing function whereas  the interband contribution $\sigma^\textrm{inter} (\ell k_\fermi)$ is a monotonically decaying function.  In our study we checked that the values $\sigma^\textrm{intra}+\sigma^\textrm{inter}$ are in most cases indistinguishable from the total conductivity. (For $N=2$ we actually see a visible but still small discrepancy.) This indicates that there are no other relevant contributions to $\sigma$ than $\sigma^\textrm{intra}$ and $\sigma^\textrm{inter}$ as defined in Eq.~\ref{sigmaintrainter}.

From the theoretical treatments of the present problem we expect $\sigma^\textrm{intra}(\ell k_\fermi)$ to contain no interband contributions. Therefore its value should be close to the Drude conductivity $\Drude = (e^2/h) \ell k_\fermi/2$, at least for large enough $\ell k_\fermi$, unless there are deviations due to for example weak (anti-)localization. For $N=2$ the interband contribution clearly follows the Drude prediction, as seen in the left panel of Fig~\ref{fig4}. Initially it surprised us that   for higher $N$ there is a clear discrepancy between the numerical results and  the Drude prediction, as can for example be seen in the right panel of Fig~\ref{fig4}. Subsequently we have understood the reason for this and hope to discuss it in detail in upcoming work. Like any semiclassical approach the Boltzmann approach, and in particular the derivation of the Drude conductivity, has as a prerequisite  a unique, well defined quasiparticle peak. For higher $N$ this condition is approximately met at low $\ell k_\fermi$, but no longer at higher $\ell k_\fermi$. Interestingly, this means that if the Boltzmann approach can at all be applied to multilayer hamiltonians, then only close to the Dirac point.  

The results in  Fig.~\ref{fig3} show  that although the artificial potential is rather different from the ordinary potential of point-like impurities, the two potentials give very similar results. The discrepancy between them for higher $N$ at large enough $\ell k_\fermi$ should have the same reasons as discussed in the previous paragraph. When the Drude picture fails, we no longer can expect our translation between $k_\fermi$ and $\ell k_\fermi$ to be correct.

We now come to what the numerical results imply for the choice of Ansatz. In the case of $N$ larger than $2$, the analytical approaches disagree and show even qualitative differences. For example, the Boltzmann approach with the Ansatz AA (equivalent to the Boltzmann equation derived with a Liouville equation approach) predicts a conductivity minimum that decreases with $N$ whereas an increase with $N$ is observed in the other considered approaches. When comparing the numerical results obtained from the Kubo formula with the analytical approaches, we realize that the agreement is mainly on the qualitative level. However, the qualitative features appear to be distinct enough for us to be able to rule out approaches.  Fig.~\ref{fig3} reveals the numerical and analytical conductivity curves for different values of $N$. As the numerical results clearly point at the conductivity minimum increasing with increasing $N$ we would rule out the Boltzmann approach with the Ansatz AA and have therefore left it out in Fig.~\ref{fig4} not to overload the figure. As the numerical results lie closer to the prediction provided by the Boltzmann conductivity with the Ansatz SKBA, we only include these and also leave out the results from the Boltzmann conductivity with the Ansatz GKBA.

\begin{figure}
 \begin{center}
 \includegraphics[width=\textwidth]{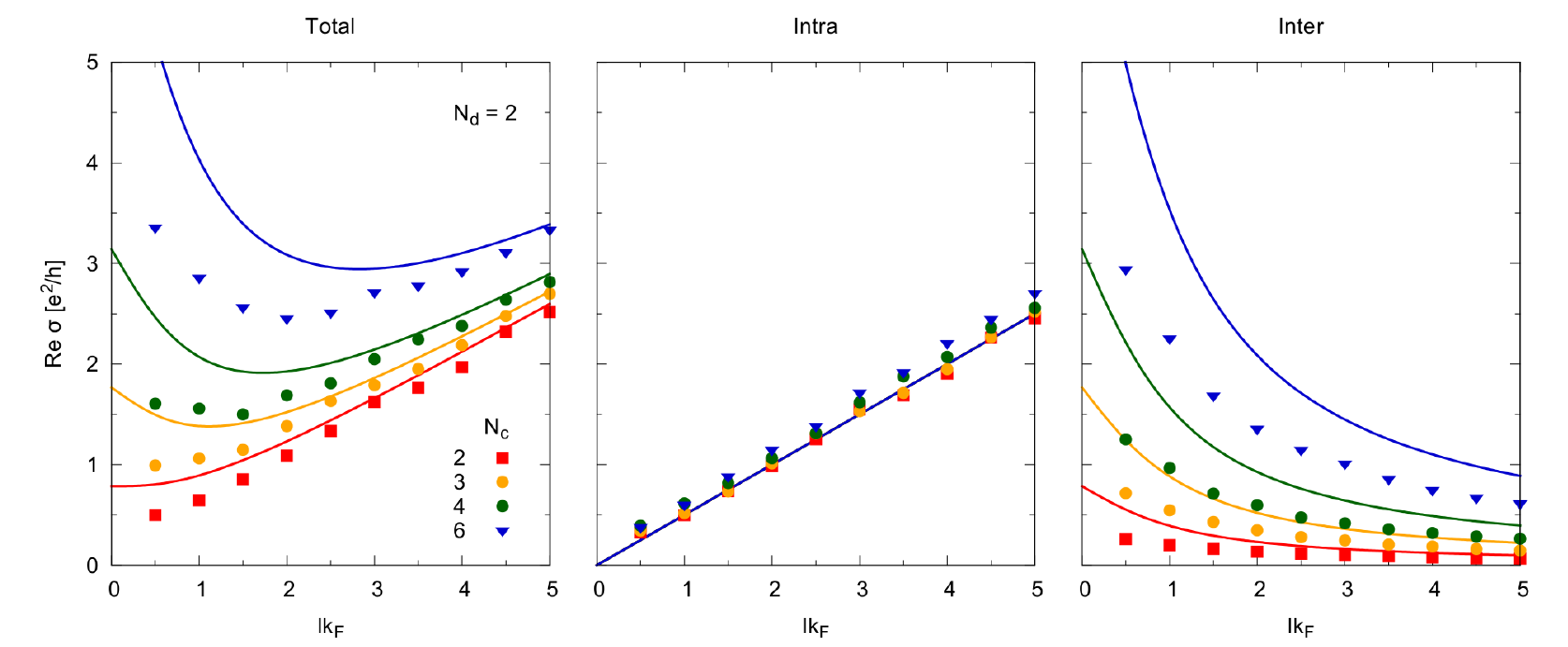}
 \caption{\label{fig5}   The total Kubo conductivity (left panel) and its components $\sigma^\textrm{intra}$  (middle panel) and $\sigma^\textrm{inter}$ (right panel) computed for the chiral hamiltonian with $\Ndisp=2$ but with chiral winding number $\Nchiral$ varied. The solid curves represent the corresponding analytical results. Note that when $\Ndisp=2$, the principle value terms do not play a role and all approaches agree, as shown in Fig~\ref{fig1}. Furthermore,  for $\Ndisp=2$ the artificial potential and the ordinary potential are identical.  The results for different $\Nchiral$ are easy to compare on the same footing since the validity of condition \eqref{e:condition3} is independent of  $k_\fermi$ and $\Nchiral$.  The parameters used are $\cD = 0.4, g_{\rm T} = 12, q_{\lambda} = 12$.}
\end{center}
\end{figure}

For our methodological studies, we have also considered the problem of fixing the power of the dispersion  in Hamiltonian \eref{NdNc} to be quadratic ($\Ndisp=2$) and only varying the chiral winding number $\Nchiral$.  This artificial model helps us to distinguish between how the chiral winding number influences the conductivity from how the power of dispersion does so. The latter factor can also be thought of as the influence of the density states, as increasing $\Ndisp$ leads to an increased density of states at the Dirac point (whereas far away from the Dirac point there is the opposite effect). Furthermore, since all the by us considered analytical approaches agree for a quadratic dispersion this setting is particularly good  for obtaining an impression of what kind of qualitative and quantitative agreement we in general can expect between the analytical and numerical approaches.
 As seen in Fig~\ref{fig5}, the qualitative agreement is good.
In particular, the local maximum of conductivity at $\ell k_\fermi =0$ increases with $\Nchiral$ in agreement with the theoretical prediction in Fig~\ref{fig1}. However, the quantitative agreement close to the Dirac point becomes increasingly unsatisfactory with increasing $\Nchiral$. The rightmost panel in Fig~\ref{fig5} shows that  the quantitative discrepancy between the numerical and analytical results observed in the total conductivity is found in the interband component.  For the intraband component, on the other hand,  the quantitative agreement is very convincing. In particular,  
 we do not find any dependence  on $\Nchiral$ in the intraband conductivity. 
We mention that we  observe the interband component to a convincing degree to obey a $\Nchiral^2$ scaling.  

For both considered problems we can conclude that the analytical approaches agree qualitatively with the behavior in the numerical results, but quantitatively the agreement is still unsatisfactory close to the Dirac point.

\section{Conclusion}\label{s:conclusion}
\noindent
In this paper we have studied the conductivity of chiral particles within the two-band model suitable for the description of low-energy electron excitations in idealized rhombohedrally stacked multilayer graphene. Our aim has not been to capture properties of a realistic graphene sample, but to use a clean, idealized setting for calculating the conductivity in the presence of chiral interband effects, and use this idealized setting for comparing analytical approaches against each other. To compute the conductivity we have employed the finite-size Kubo formula and performed an exact diagonalization of the total Hamiltonian consisting of the mentioned chiral kinetic Hamiltonian and a point-like impurity potential. Among the analytical approaches we both presented a heuristic derivation based on the Kubo formula as well as studied a pseudospin-coherent generalized Boltzmann equation. In the latter we stressed the leading role of the chiral winding number $\Nchiral$ in the band-coherent conductivity and showed that the choice of Ansatz has a huge impact on the results  close to the Dirac point, see Figures~\ref{fig1},~\ref{fig2}.

In the numerical analysis we saw that the conductivity minimum increases with the layer number  $N$, see Figure~\ref{fig3}. Comparison with the analytical results rules out the band-coherent Boltzmann equation derived with a Green's function approach using the Ansatz AA (the one with $\nu=-1$ in the principal value collision integral terms, see Figure~\ref{fig2}) which predicts the opposite behavior. The band-coherent Boltzmann equation is equivalent to the one derived within the Liouville equation approach, which is the simplest and most popular approach of deriving generalized Boltzmann equations in the spin-orbit interaction context. The conclusion made above therefore rules out the Liouville equation approach as well. This approach seems to be unsuitable in general when band-coherent effects become important, although it might be acceptable as long as the principle value collision integral terms do not play a role. The evidence against the Liouville equation approach for the derivation 
of the principal value terms in the collision integral is one of the main and most surpising results of the present paper, especially if one considers the simplicity and apparent self-sufficiency of the Liouville-like derivation.

On the constructive side our hope was to single out one approach for deriving band-coherent Boltzmann equations. The Green's function derivation with the Ansatzes GKBA ($\nu=1$) and SKBA ($\nu=0$) result in analytical predictions closer to the numerical results than the mentioned derivation using the Liouville equation. They predict correctly a conductivity minimum which increases with the number of layers, see Fig~\ref{fig3}. The best results have been obtained within the approach using the Ansatz SKBA.  In the corresponding collision integral the principal value terms are absent by exact cancellations.  The discarding of the competitor GKBA would also entail the  discarding of the Liouville-type derivation with the Markov approximation done not in the interaction picture but in the Schr\"odinger picture, see Appendix~\ref{s:liouville}.  Support for the Ansatz SKBA suggests that principal value terms are not a part of the correct band-coherent generalization of the Boltzmann collision integral. In an approach were they are present one does best in neglecting them. One example of this is the appendix in ref.~\onlinecite{ShytovPRB2006}, were a Green's function approach with the GKBA is used. The principal value terms found in the derivation are neglected, without detriment to the end result. 

%The favoring of the Ansatz SKBA in front of GKBA comes with the following caveat. In our treatment we acknowledged that real parts of the self-energy terms should probably not be seen as a part of the collision integral but instead be transferred to the drift side, were they renormalize the free drift. (Only in a Green's function approach but not in a Liouville type of approach does such a separation seem to be natural.) In the by us considered problem it appeared that the real parts are zero. However, if for some reason our treatment of the real parts would not be correct, one might hope that accounting for this  renormalization would lead to an improved quantitative agreement.  The outcome might change the order in which the Ansatzes seem to be suitable. It seems very unlikely, though,  that the Ansatz AA would come back into the game.    

One interesting observation was made when we decoupled the chiral winding number $\Nchiral$ from the power $\Ndisp$ of the dispersion, and kept the latter fixed at $\Ndisp=2$, in which case all analytical approaches agree. As seen  in the leftmost panel of Figure~\ref{fig5}   the total conductivity shoots up  when $\ell k_\fermi $ approaches the neutrality point. Moreover, the value at $\ell k_\fermi$ increase  with the winding number $\Nchiral$, as predicted by our analytical expression. For a given $\Nchiral$ the value at the Dirac point is larger than in the case when $\Ndisp=\Nchiral$, shown in  Figure~\ref{fig3}. This supports the analytical prediction that an  increase of  the winding number $\Nchiral$ increases the conductivity close to the Dirac point, whereas the increase of the power of dispersion $\Ndisp$ has the opposite effect. The latter effect  is somewhat surprising considering that increasing $\Ndisp$ leads to an increased density of states at the Dirac point.  In the multilayer case $\Nchiral =\Ndisp = N$ these two effects partially cancel each other, with the net effect still being an increase of the conductivity at the Dirac point with increasing $N$. 

%%% Next section is logically independent.

At higher values of $\ell k_\fermi$, the numerically calculated conductivity for higher layer number $N$ showed a feature that we did not expect from the analytical treatment. The total conductivity is found to be significantly below the Drude conductivity at higher electron densities, although one would imagine that one is safely in the Boltzmann regime $ \ell k_\fermi \gg 1$ far away from the Dirac point.  We have found that this discrepancy between the numerical Kubo and analytical Drude predictions is due to the lack of a unique quasiparticle peak and can be eliminated if the scattering potential has not a point-like structure. The problem is well known for the delta-functional scattering potential in the two-dimensional case. (See e.g. the review  \cite{PeresRMP2010}.) The Born approximation tends to underestimate the scattering cross section in this case. Since this issue does not seem to be  related to interband coherence and to  chirality but  rather to the properties of spectral functions of excitations in a given system, we postpone the discussion to subsequent work dealing with the electron transport of chiral particles far from the neutrality point. 

To conclude, we have shown that the band-coherent Boltzmann approach has a potential of being a theory of conductivity of chiral particles close to the neutrality point, even though this approach seems to be formally invalid in this region. For the derivation of Boltzmann-like band-coherent kinetic equations the Liouville equation approach should be used with great precaution. In particular, it turns out to be unsuitable for the derivation of the principal value terms in the collision integral. However, it  can probably be used for the qualitative analysis of the band-coherent conductivity as long as these terms play no role.

\begin{acknowledgments}
The authors wish to thank Pavel Lipavsk{\' y}, Klaus Morawetz and John Schliemann for useful discussions. This work was supported by the Program of „Employment of Newly Graduated Doctors of Science for Scientific Excellence“ (grant number CZ.1.07/2.3.00/30.0009) co-financed from European Social Fund and the state budget of the Czech Republic. The access to computing and storage facilities owned by parties and projects contributing to the National Grid Infrastructure MetaCentrum, provided under the programme "Projects of Large Infrastructure for Research, Development, and Innovations" (LM2010005) is also highly appreciated.
\end{acknowledgments}

\appendix 
\section{Calculations used for the band-coherent Boltzmann treatment}\label{s:boltzcalc}
\noindent
\be{noname-28}
\Lambda_k &=& \frac{2 W_0}{(2\pi)^2 \ben} \int_0^\infty k\pr\der k\pr \, \frac{1}{1-\left (\frac{\ben\pr}{\ben}\right)^2}
=\left [ x \equiv \frac{k}{k_\fermi}, \quad z\equiv \frac{k\pr}{k}, \quad w\equiv z^2\right] =
 \frac{2 W_0 k^2}{(2\pi)^2 \ben} \int_0^\infty z \der z \, \frac{1}{1-z^{2\Ndisp}} =
 \nonumber \\
 &=& 
 \frac{\Ndisp}{\pi \ttr} x^{2-\Ndisp} \int_0^\infty  \der w \, \frac{1}{1-w^{\Ndisp}} 
  \nonumber \\
 &=& 
 \frac{\cot \frac{\pi}{\Ndisp}}{\ttr} \left(\frac{k}{k_\fermi}\right)^{2-\Ndisp}
 \, .
\ee

 With  
\be{noname-29}
\frac{W_0 k^2}{\ben} = \frac{W_0 k_\fermi^2}{\ben_\fermi} \left (\frac{k}{k_\fermi}\right)^{N-2} =
 \frac{4\pi N}{\ttr} \left (\frac{k}{k_\fermi}\right)^{N-2} 
\ee
we have in equation \eref{sigma0sigma}
\be{noname-30}
\begin{array}{l}
\int_{k_\fermi}^\infty   \der k \, \frac{\ben }{k} \frac{\IC}{(2 \ben-\Ansatzindex\Lambda)^2 +\IC^2} \quad =  
\left [ x\equiv \frac{k}{k_\fermi} , \beta\equiv 2 \ben_\fermi \ttr, c \equiv \Ansatzindex \cot \tfrac{\pi}{N}\right ]= 
  \frac{1}{2} 
    \int_1^\infty \der x \frac{\beta x }{(\beta x^{ \Ndisp} -c x^{2-\Ndisp})^2+x^{4-2 \Ndisp}}  =
    \frac{1}{2} \int _1^\infty \der x \frac{\beta x^{2 \Ndisp-3}}{(\beta x^{2\Ndisp - 2}-c)^2+1} = 
    \nonumber 
    \\
   =\left [ y =\beta x^{2 \Ndisp - 2}\right] = \frac{1}{2}
    \int_\beta^\infty \frac{\der y}{2\Ndisp -2} \frac{1}{(y-c)^2+1} =
    \frac{1}{4(\Ndisp - 1)}  \left [ \arctan y\right ]_\beta^\infty     
 = \frac{1}{4(\Ndisp-1)}\left (\frac{\pi}{2}-\arctan \left(2 \ben_\fermi \ttr-\Ansatzindex\cot \frac{\pi}{\Ndisp}\right)\right) \, .  
     \end{array} 
\ee

\section{Derivation of the collision integral from the quantum Liouville equation}\label{s:liouville}
\noindent
The simplest derivation of a collision integral is probably the  one in which the quantum Liouville equation is iterated to second order in the interaction. Let $\Htot=\Ho+\Hi$ where $\Hi$ is  an  
interaction switched on at a time $t_0$ in the remote past.  The quantum Liouville equation for the density matrix $\rho_{s_1s_2}(t,x_1,x_2)$   in  the interaction picture is
\be{Lioint}
i\partial_t \rho^\textrm{I}=[\Hi\INT (t),\rho^\textrm{I}]
\ee
 with 
$A\INT(t)~=~e^{i\Ho t}A(0) e^{-i\Ho t}=\evu_0^\dagger(t)A(0)\evu_0(t) $. This is easily integrated to  give
\be{intLioint}
\rhoI(t)~=~\rhoI(t_0)-i\int_{t_0}^t \der t\pr [\Hi\INT(t\pr),\rhoI (t\pr)]  
\ee
which, when inserted back into  \eref{Lioint}, yields 
\be{VVV}
\partial_t \rho^\textrm{I}&=&-i[\Hi\INT  (t),\rho^\textrm{I}(t_0)]  -\int_{t_0}^t \der t\pr [\Hi\INT(t), [\Hi\INT(t\pr),\rhoI (t\pr)]]~ =
\nonumber \\
&=&
-i[\Hi\INT  (t),\rho^\textrm{I}(t_0)]  -\int_{t_0}^t \der t\pr [\Hi\INT(t), [\Hi\INT(t\pr),\rhoI (t)]] 
-i\int_{t_0}^t \int_{t\pr} ^{t} \der t\pr [\Hi\INT(t), [\Hi\INT(t\pr), [\Hi\INT(t^{\prime\prime}),\rhoI (t^{\prime\prime})]] 
\ee
after a first and second iteration, respectively (the second time with the interval of integration $(t_0,t)$ replaced by $(t\pr,t)$ ).
 So far the equations are exact.  The Born approximation allows us to get  a closed equation for $\rho$ at time $t$ to second order in the interaction $\Hi$. To this end we remove the last term in the second row. Equivalent to this is to replace in the last term of  the  first row,  the full evolution  with the  free evolution, \ie  let  
$\rhoI(t\pr)=\rhoI(t)$, which   
has the appearance of a Markov approximation. 
Back in the \sch{}picture the assumption of free evolution reads
\be{correctmarkov}
\rho(t\pr )~=~ \evu_0(t\pr,t)\, \rho(t)\,\evu_0^\dagger (t\pr,t)~=~e^{ -i \Ho (t\pr,t)}\, \rho (t)\, e^{i \Ho (t\pr ,t)} \, 
\ee
 and after 
reorganizing evolution operators one obtains the kinetic equation in the \sch picture,
 \be{noname-05} 
\partial_t\rho (t) + i[\Ho,\rho(t) ]= 
-i[\Hi, e^{-i\Ho (t-t_0)}\rho(t_0)e^{i\Ho(t-t_0)} ] -
  \int_0^{t-t_0}  \der \tau [\Hi,[e^{-i\Ho \tau }\Hi e^{i\Ho\tau},\rho(t)]   ] .
   \ee
 So far, this kinetic equation is locally time reversible (globally not, since we switched on the interaction).
To capture the decoherence due to other processes (phonons etc.) we do not want the state to depend on correlations in the remote past. Therefore we include a factor $e^{-\eta\tau}$ in the integral to impose this loss of memory. This introduces time irreversibility. This factor also regularizes the integral and allows us to send $t_0\rightarrow -\infty$.  In translating   
the evolution operators 
into Green's functions, the last term can be written  as
\be{jme}
\JC[\rho]&=&-\int \frac{\der \omega}{2\pi}[\Hi,[\GoR\Hi\GoA,\rho]],
\ee
where we anticipated that the last term will become the collision integral $\JC$.

Another source of irreversibility comes with the impurity averaging procedure, a coarse graining that also captures the decoherence due to  
phonons, for example.  Then $\Hi$ in the first term becomes just a number $\sim V(r=0) $  and the commutator  vanishes.\footnote{
In the derivation of master equations for quantum dots this term 
vanishes for another reason, see \eg ref.~\cite{Timm-2008}. For $t\leq t_0$ the state is a simple product state $\rho=\rho_\textrm{dot}\otimes \rho^\textrm{eq}_\textrm{lead}$ with the leads assumed to be in equilibirum. Since $ \rho^\textrm{eq}_\textrm{lead}$ has a definite particle number whereas the hopping interaction $\Hi$ changes the particle number, tracing over the leads kills this term. 
}   
For the terms linear in $\nimp$ one finds for example (summation over repeated indices implicit) 
\be{noname-31}
(\Hi \GoR \Hi \GoA \rho)_{k k\pr } 
\longrightarrow
\nimp\delta (k-k_1+k_1-k_2) \singleimp_{kk_1}\GoR_{k_1} \singleimp_{k_1k_2}\GoA_{k_2} \rho_{k_2k\pr}= \SR_{k}\GoA_k\rho_{kk\pr},
\ee
where we introduced the retarded self-energy $\SR_k=\nimp(\singleimp \GoR\singleimp)_k$. 
However, in a term like 
\be{noname-32}
(\Hi \rho \GoR \Hi \GoA)_{kk\pr}
\longrightarrow 
\nimp
\delta (k-k_1+k_2-k\pr) \singleimp_{kk_1} \rho_{k_1k_2}\GoR_{k_2}\singleimp_{k_2k\pr}\GoA_{k\pr} 
\ee  
 the delta function seems to offer no simplification at all.

At this point we can attain further simplification if we say that in the collision integral we are not interested in any contributions which have to do with non-diagonality in momentum.  This actually amounts to saying that we are not interested in any gradient expansion corrections to the collision integral: 
\be{noname-33}
(\Hi \rho \GoR \Hi \GoA)_{kk} 
\longrightarrow \nimp  \singleimp_{kk\pr} \rho_{k\pr} \GoR_{k\pr}\singleimp_{k\pr k} \GoA_k\, .
\ee
The collision integral can then be written as 
\be{liouvilleJ}
\JC[\rho(\vec{k},\vec{x},t) ]   ~=~  -\ki  \Wkk \int \frac{\der \omega }{(2 \pi)^2} 
(\rho_\vec{k}\GoR_\vec{k} \GoA_\vec{k\pr}+
 \GoR_\vec{k\pr}\GoA_\vec{k} \rho_\vec{k}
 -\rho_\vec{k\pr}\GoR_\vec{k\pr}\GoA_\vec{k} -
 \GoR_\vec{k} \GoA_\vec{k\pr}\rho_\vec{k\pr}
 )
\ee
with the transition matrix $\Wkk=2\pi n_\textrm{imp}|\singleimp_{\vec{k}\vec{k\pr}} |^2$ for spinless impurities. 
The collision integral \eref{liouvilleJ} is identical to the one derived with the Green's function approach \GO\AA. The collision integral will in general contain both delta function terms and principal value terms. 

Note that if one were to adopt a Markov approximation instead the Schr\"odinger pciture as $\rho(t\pr)\rightarrow \rho(t)$  (as done for example in ref.~\onlinecite{Culcer-Winkler-2007b})
rather than $\rho\INT (t\pr)\rightarrow \rho\INT (t)$ as in the Dirac picture, then the evolution operators cancel  each other out in a different way so that at the end they sit around the entire inner commutator rather than only around the inner $\Hi$, 
\be{culcerj}
\JC[\rho]~=~ -\int \frac{\der \omega}{2\pi}[\Hi,\GoR [\Hi,\rho]\GoA]\, .
\ee
This last Markov approximation is not equivalent to neglecting the $\ordo{V^3}$ term in the last line of \eref{VVV}, that is, it is not equivalent to what we understand as the Born approximation. What is a correct Markov approximation is not important to us. The base line is that we want to derive the collision integral in the  lowest Born approximation, and for this the steps leading to equation \eref{jme} seem to us to be the right ones. 

The different structure of of the double commutators in \eref{jme} and \eref{culcerj} are crucial. The collision integral \eref{culcerj} coincides now with the one derived with the Green's function approach \GO GKBA.

\section{Simple example of why one needs an Ansatz}\label{s:whyansatz}
\noindent
Suppose that by definition 
\be{noname-34}
 \int \der \omega \,   A(\omega,t,\vec{x},\vec{p})   &\equiv& a(t,\vec{x},\vec{p}) \nonumber \\
  \int \der \omega \,   B(\omega,t,\vec{x},\vec{p}) & \equiv& 1
  \ee
 for some quantum correlation functions $A$ and $B$. Although there is a simple relation between   
 $A(\omega,t,\vec{x},\vec{p}) $  and $a(t,\vec{x},\vec{p}) $, the former will typically contain much more information than the latter, and some of the information goes lost when we integrate over the variable $\omega$. 
 
Suppose there is an equation of  motion such as
\be{noname-38}
\partial _t  A(\omega,t,\vec{x},\vec{p}) +\vec{v}_\vec{p} \cdot \partial_\vec{x} A(\omega,t,\vec{x},\vec{p}) = 0\, .
\ee 
We can integrate the whole equation over $\omega$ to get rid of this variable and used the definitions to obtain 
\be{noname-35}
\partial _t  a(t,\vec{x},\vec{p}) +\vec{v}_\vec{p} \cdot \partial_\vec{x} a(t,\vec{x},\vec{p}) = 0\, ,
\ee
that is, 
a closed equation for $a(t,\vec{x},\vec{p}) $, meaning that the original unknown quantity $A$ only enters in the form of $a$.  The latter equation of motion contains of course less information than the former (some of the information was lost in the $\omega$-integration), but might be that the information in the latter equation is just about the amount of information we are interested in and that the equation is now analytically more tractable.  

However, for an equation like 
\be{exampleeq2}
\partial _t  A(\omega,t,\vec{x},\vec{p}) +\vec{v}_\vec{p} \cdot \partial_\vec{x} A(\omega,t,\vec{x},\vec{p}) =  \int_\vec{p\pr}  \Gamma_{\vec{p}\vec{p\pr}}B(\omega,t,\vec{x},\vec{p})  A(\omega,t,\vec{x},\vec{p\pr}) 
\ee 
we have for the right hand side no use of the definitions when we want to integrate the whole equation over $\omega$ to get rid of this variable. The definitions tell us only what the outcome in terms of $\omega$ independent quantities are when  $A(\omega,t,\vec{x},\vec{p})$ and $B(\omega,t,\vec{x},\vec{p})$ are standing alone or only together with quantities that do not depend on $\omega$.  There is no general identity that would give us a result of the integral
\be{noname-36}
\int \der \omega \, B(\omega,t,\vec{x},\vec{p})  A(\omega,t,\vec{x},\vec{p}).
\ee
To make progress we need more explicit information about $A(\omega,t,\vec{x},\vec{p})$ and $B(\omega,t,\vec{x},\vec{p})$ than what is contained in the definitions. For example,  we might know that we can work with
\be{exampleapprox}
A(\omega,t,\vec{x},\vec{p})   & = & a(t,\vec{x},\vec{p}) \delta (\omega -\erg_p)  \nonumber \\
B(\omega,t,\vec{x},\vec{p}) &  = &  \delta (\omega -\erg_p)\, .
\ee 
These relations are consistent with the definitions but provide us with more information that do the definitions. 
We might have arrived at these simple expressions by arguing that they are sensible approximations in our context. These relations might be the result of our simplifying the information contained in $A$ and $B$. With these last equations we can integrate equation \eref{exampleeq2} over $\omega$ and fully get rid of any $\omega$-dependence, namely 
 \be{exampleeq3}
\partial _t  a(t,\vec{x},\vec{p}) +\vec{v}_\vec{p} \cdot \partial_\vec{x} a(t,\vec{x},\vec{p}) =  \int_\vec{p\pr} \delta (\erg_p-\erg_{p\pr})  \Gamma_{\vec{p}\vec{p\pr}}  a(t,\vec{x},\vec{p\pr}) \, .
\ee 
This is a closed eqution for $a$ without any explicit appearance of $A$ and $B$ or of other $\omega$ dependent quantities. Without something like the expressions \eref{exampleapprox} we would have not been able to arrive at a closed equation such as \eref{exampleeq3}.

%\bibliography{graphene.bib,Maxim2Janik.bib,Kailasvuori_Bibtexfile.bib}

\end{document}